\documentclass[9.5pt,journal,compsoc]{IEEEtran}

\ifCLASSOPTIONcompsoc
  \usepackage[nocompress]{cite}
\else
  \usepackage{cite}
\fi

\usepackage{balance}
\usepackage{verbatim}
\usepackage{graphicx}
\usepackage{booktabs,chemformula}
\usepackage{multirow}
\usepackage{url}
\usepackage{amsmath} 
\usepackage{tcolorbox}
\usepackage{xspace}

\newcommand{\paper}{paper\xspace}

\usepackage{xcolor}

\PackageWarning{TODO:}{X}
\PackageWarning{TODO:}{X\%}

\newcommand{\etal}{\hbox{\emph{et al.}}\xspace}
\newcommand{\eg}{\hbox{\emph{e.g.,}}\xspace}
\newcommand{\ie}{\hbox{\emph{i.e.}}\xspace}

\newcommand{\authorshipRec}{\emph{AuthorshipRec}\xspace}
\newcommand{\reviewRec}{\emph{RevOwnRec}\xspace}

\newcommand{\cHRev}{\emph{cHRev}\xspace}
\newcommand{\learnRec}{\emph{LearnRec}\xspace}
\newcommand{\retentionRec}{\emph{RetentionRec}\xspace}
\newcommand{\turnoverRec}{\emph{TurnoverRec}\xspace}
\newcommand{\sofiaEhsan}{\emph{Sofia}\xspace}

\newcommand{\whodo}{\emph{WhoDo}\xspace}

\newcommand{\sofiaTwo}{\emph{SofiaWL}\xspace}

\newcommand{\expertise}{\textit{Expertise}\xspace}
\newcommand{\dExpertise}{$\Delta$Expertise\xspace}

\newcommand{\workload}{\textit{GiniWork}\xspace}
\newcommand{\dWorkload}{$\Delta$GiniWork\xspace}

\newcommand{\far}{\textit{FaR}\xspace}
\newcommand{\dFar}{$\Delta$FaR\xspace}

\newif\ifIsThesis
\IsThesisfalse

\usepackage{array}

\begin{document}
%\title{Code review recommendation to balance review workload, mitigate knowledge loss from turnover, and to ensure expert review}
\title{Factoring Expertise, Workload, and Turnover into Code Review Recommendation}
%\title{Code review recommendation to find experts, balance review workload, and mitigate knowledge loss from turnover}
%\title{Reviewer recommendation to balance workload and reduce knowledge loss and ensuring expertise during code review}
%\title{Code Review Recommendation to Mitigate Turnover Induced Knowledge Loss: Balancing Expertise, Workload, and Knowledge Spreading}

\author{Fahimeh Hajari, Samaneh Malmir, Ehsan Mirsaeedi, 
        and Peter~C. Rigby~\IEEEmembership{}% <-this % stops a space
\IEEEcompsocitemizethanks{\IEEEcompsocthanksitem M. F Hajari, S. Malmir, E Mirsaeedi, and P. C. Rigby are with the Department of Computer Science and Software Engineering, Concordia University, Montr{\'e}al, Qu{\'e}bec, Canada.\protect\\
E-mail: f\_hajari@encs.concordia.ca, samaneh.malmir@concordia.ca, s\_irsaee@encs.concordia.ca, peter.rigby@concordia.ca}% <-this % stops an unwanted space
% \thanks{Manuscript received April 19, 2005; revised August 26, 2015.}
}

\markboth{Transactions on Software Engineering,~Vol.~X, No.~Y, November~2023}%
{Shell \MakeLowercase{\textit{Hajari \etal}}: Reviewer recommendation to balance workload and reduce knowledge loss and ensuring expertise during code review}

\IEEEtitleabstractindextext{
    %\justifying
    
Developer turnover is inevitable on software projects and leads to knowledge loss, a reduction in productivity, and an increase in defects. Mitigation strategies to deal with turnover tend to disrupt and increase workloads for developers. In this work, we suggest that through code review recommendation we can distribute knowledge and mitigate turnover while more evenly distributing review workload. 

We conduct historical analyses to understand the natural concentration of review workload and the degree of knowledge spreading that is inherent in code review.
%Code reviews natural spread knowledge, and knowledgeable reviewers reduce the files at risk to turnover by 47.2 percentage points. 
Even though review workload is highly concentrated, we show that code review natural spreads knowledge thereby reducing the files at risk to turnover.

Using simulation, we evaluate existing code review recommenders and develop novel recommenders to understand their impact on the level of expertise during review, the workload of reviewers, and the files at risk to turnover. Our simulations use seeded random replacement of reviewers to allow us to compare the reviewer recommenders without the confounding variation of different reviewers being replaced for each recommender. 
%Our simulation involves replacing the same set of seeded random reviewers to control natural variation and exclusively evaluate how each recommender would have performed. 

We find that prior work that assigns reviewers based on file ownership concentrates knowledge on a small group of core developers increasing the risk of knowledge loss from turnover. 
Recent work, WhoDo, that considers developer workload, assigns developers that are not sufficiently committed to the project and we see an increase in files at risk to turnover.
We propose learning and retention aware review recommenders that when combined 
are effective at reducing the risk of turnover, but they unacceptably reduce the overall expertise during reviews.

Combining recommenders, we develop the \sofiaTwo recommender that suggests experts with low active review workload when none of the files under review are known by only one developer. In contrast, when knowledge is concentrated on one developer, it sends the review to other reviewers to spread knowledge. For the projects we study, we are able to globally increase expertise during reviews, $+3\%$, reduce workload concentration, $-12\%$, and reduce the files at risk, $-28\%$. 

We make our scripts and data available in our replication package~\cite{replication}. Developers can optimize for a particular outcome measure based on the needs of their project, or use our GitHub bot to automatically balance the outcomes~\cite{Mirsaeedi2020ICSE}. 

%\todo{Our final contribution is to evaluate the our proposed recommenders and those from the literature in the context of suggesting learner developers. Unlike a busy expert, learners should have less of an impact on overall workload and should naturally reduce the files at risk.}

%
%\todo{remove? \sofiaTwo is integrated into GitHub pull requests allowing developers to select an appropriate expert or ``learner" based on the context of the review and the developers' workloads. We release the \sofiaTwo bot as well as the code and data for replication purposes.}
    
}

\maketitle

\IEEEdisplaynontitleabstractindextext

\IEEEpeerreviewmaketitle

\maketitle

\section{Introduction}
%\point{Turnover is bad, cite papers}
Turnover on software projects is frequent and inevitable and leads to the loss of knowledge when developers leave a project~\cite{bao2017will,Rigby2016ICSE}. Turnover incurs substantial economic cost in recruiting and training new employees~\cite{pekala2001holding,mockus2010organizational}, it reduces the productivity of development teams~\cite{mockus2010organizational,izquierdo2009using}, it leads to the loss of critical tacit knowledge~\cite{mockus2009succession,mockus2010organizational,huselid1995impact}, and has been shown to increase the number of defects in a product and reduce overall product quality~\cite{mockus2009succession,foucault2015impact,mockus2010organizational}. 

%\point{Mitigation is expensive}
Recent works have tried to mitigate the adverse impact of turnover through increasing knowledge retention by predicting leavers~\cite{lin2017developer,bao2017will,constantinou2017empirical}, planning for succession~\cite{stovel2002voluntary,pee2014mitigating,mockus2009succession}, documenting knowledge, and persisting knowledge on StackOverflow and other internal QA forums~\cite{rashid2017exploring,pee2014mitigating}. However, these mitigation practices often require organizational changes and additional developer effort especially by those who are expert enough to answer questions and write documentation~\cite{rashid2017exploring}.

%\point{code review distribute knowledge}
In this work, we show that code review can mitigate turnover risk because it naturally distributes knowledge by exposing developers to code they have not authored during reviews. Prior work interviewed developers and showed that code review is an opportunity for learning and it plays a vital role in distributing knowledge~\cite{bacchelli2013expectations,greiler2016code,bosu2016process}. 
Furthermore, studies have quantified the potential knowledge gained during code review~\cite{rigby2012contemporary,sadowski2018modern} and shown that developers review code in modules they have not modified~\cite{thongtanunam2016revisiting}.
In contrast to other turnover mitigation strategies, code review is a common and well-established practice in teams that does not require teams and individuals to alter their current workflow. 

%\point{recommenders}
In this work, we enhance code review's inherent knowledge sharing potential by developing review recommenders to distribute knowledge and use simulations to show that they mitigate turnover risk.
%
%\point{existing ones ignore knowledge sharing and their suggestions are obvious} 
In contrast, existing review recommenders are solely focused on finding expert reviewers and disregard the role of code review in distributing knowledge among developers, \eg~\cite{xia2015should,yu2016reviewer,rahman2016correct}. These recommenders result in expertise concentration because the evaluation benchmark measures how many of the actual developers who performed the review were recommended. Interviewed developers state that these recommenders suggest obvious candidates and do not provide additional value~\cite{kovalenko2018does}.

The second goal of this work is to make recommenders aware of developer workload. Core developers tend to be both committed and expert~\cite{rigby2014peer,kononenko2016code}, and are ideal review candidates to increase expertise and reduce the files at risk to turnover. However, core developers already have a high workload and a recommender that increases the workload of the top reviewers would be detrimental to the project. Recent work at Microsoft introduced the WhoDo recommender that factors developer review workload to ensure that expert developers are not overwhelmed by reviews~\cite{asthana2019whodo}. We reimplement the WhoDo recommender and evaluate it on open source projects. We then develop a recommender, \sofiaTwo, that suggest experts that currently have low active workloads and learners who spread knowledge. 

To evaluate recommenders we introduce three outcomes \expertise, \workload, and \far. The first outcome ensures that expertise remains high for finding defects during review, \ie the reviewers know about the files under review. The second, ensures that the top reviewers are not unreasonably overworked due to always being the top recommendation, \ie the change in area under the review workload distribution curve or Gini coefficient. 
%The third, ensures that average reviewer does not receive an unreasonable increase in review workload. 
The third outcome measures the number of files that are at risk to turnover, \far, to ensure that knowledge is adequately distributed during review, \ie how many people know about each file. We run simulations on the historical reviews of five large projects to understand how recommenders affect each outcome. For completeness, we also calculate Mean Reciprocal Rank, MRR, to understand how well each recommender predicts the developers who actually performed the review. 
%\point{how we design our novel recommender}
%We then develop a review recommender that is knowledge and developer retention potential aware. We are able to balance the need to have expertise, while reducing the workload on the core team, and providing learning opportunities to developers to reduce the risk of turnover. The following research questions elaborate on our the evaluated recommenders and outcomes.

\subsection{Paper Extension and Research Questions}

This work is an extension over Mirsaeedi and Rigby's~\cite{Mirsaeedi2020ICSE} previous work with the novel focus on code review workload. We adapt the economic measure of the Gini coefficient and the Lorenz curve to understand the concentration of reviewer workload (rather than the concentration of money) in methodology Section~\ref{secSimulation}. We then re-evaluate the existing recommenders that are not aware of workload in Research Questions 2 and 3 to understand if they distribute or concentration reviewer workload. Research Questions 4, 5, and 6 are entirely new to this paper: we empirically describe reviewer workload, evaluate workload aware recommendation, and ensure that we continue to balance \expertise and reduce \far. We provide a detailed discussion of reviewer workload and workload measures in Section~\ref{secRelatedWork}.

We divide our research questions into two types: those that are purely based on a {\it historical analysis} and evaluate the actual state of the software project, and those that use {\it simulation} to evaluate each recommender on our outcome measures. As we discuss in Section~\ref{secSimulation}, our simulations use seeded random replacement of reviewers to allow us to compare the techniques without the confounding factor of different reviewers being replaced for each recommender. 

We provide evidence related to six research questions:

\begin{itemize}

\item RQ1. 
%Review and Turnover (Historical Analysis): 
What is the reduction in files at risk to turnover when both authors and reviewers are considered knowledgeable of the files in a pull request?

\item RQ2. 
%Ownership Aware (Simulation):
How does recommending reviewers based on the files they have authored and reviewed in the past impact the number of files at risk to turnover?

\item RQ3. 
%Turnover Aware (Simulation): 
Can we reduce the number of files at risk to turnover by developing learning and retention aware review recommenders?

\item RQ4. 
%Review Workload (Historical Analysis): 
How is the review workload distributed across developers?

\item RQ5.
%Workload Aware (Simulation): 
WhoDo is designed to be workload aware, but what is its impact on reviewer expertise and the files at risk to turnover?
%can it also balance \expertise, \workload, and \far?

\item RQ6. 
%Ownership, Turnover, and Workload Aware (Simulation): 
Can we combine the recommenders to balance reviewer expertise, workload, and the files at risk to turnover? 
%\expertise, \workload, and \far?

\end{itemize}

This \paper is organized as follows.
%remove for thePaper
%In Section~\ref{secSurvey}, we conduct a board survey of the literature on turnover, code review, and reviewer recommenders. 
In Section~\ref{secBackgroundDefinitions}, we provide the study background as well as defining our measures, review recommenders, scoring functions, and simulation methodology. In Section~\ref{secData}, we describe the projects under study and project data. In Section~\ref{secResults}, we present results for each of our research questions. 
%In Section~\ref{secSophia}, we describe the \sofiaEhsan bot which integrates into GitHub pull request. 
In Section~\ref{secThreats}, we discuss threats to validity. In Sections~\ref{secRelatedWork} 
%add back for thePaper: 
and \ref{secConclusion}, 
we discuss our findings in the context of the existing literature, future work, and conclude the \paper.

\section{Background and Definitions}
\label{secBackgroundDefinitions}

%easy to cut
In this section, we introduce the background on ownership, review recommenders, reviewer workload, and knowledge loss and show the manner in which each has been quantified in the past and suggest novel measures and outcomes. We will subsequently use these measures as the basis on which to expand reviewer recommendation in a scoring function that will also be workload and knowledge aware. 

\subsection{The Ownership Recommenders}
\label{secOwnership}

The influence of code ownership on code quality has been extensively investigated in the literature\cite{bird2011don,rahman2011ownership,foucault2015impact,thongtanunam2016revisiting}. Ownership is a human factor that helps with finding knowledgeable developers that can be accountable for a particular part of code or task~\cite{mockus2002expertise}. \textit{Developer Recommenders} use ownership to automatically assign tasks to experts~\cite{kagdi2008can}. Researchers have used a wide range of granularity, from lines~\cite{fritz2007does, girba2005developers,rahman2011ownership} to modules~\cite{bird2011don}, to estimate ownership of developers. 
Studies on code review find that code owners are usually selected to review changes~\cite{sadowski2018modern,bacchelli2013expectations,greiler2016code}. In this work we develop two simple scoring functions for review recommendation based on ownership.

\textbf{\authorshipRec.} Bird~\etal~\cite{bird2011don} define code ownership for a developer in a module as the ratio of commits the developer has made relative to the total commits made to that component. Our \authorshipRec scores a developer, $D$, as a candidate reviewer based on the number of commits he or she has made to the files under review, $R$, divided by the total number of commits made to these files. 
%Given the function $Commit(D, F)$ that returns the number of commits developer $D$ has made on file $F$, the $CommitOwnership(C, R)$ of candidate $C$ given review $R$ is defined as

$\text{AuthorshipRec}(D, R) =$
\begin{equation}
 \frac{\text{CommitsForFilesUnderReview}(D,R)}{\sum_{d}^{Devs}\text{CommitsForFilesUnderReview}(d,R)}
\label{eqAuthorshipOwnership}
\end{equation}

\textbf{\reviewRec.} Thongtanunam \etal~\cite{thongtanunam2016revisiting} devise a review aware ownership metric based on the files that a developer has reviewed. Intuitively, reviewers who have reviewed the changed files or modules in the past, will be good candidate reviewers. To recommend reviewers, we score the number of times a candidate has reviewed the files in the past divided by the total number of times the files have been reviewed.

$\text{RevOwnRec}(D, R) =$
\begin{equation}
 \frac{\text{ReviewsOfFilesUnderReview}(D,R)}{\sum_{d}^{Devs}\text{ReviewsOfFilesUnderReview}(d,R)}
\label{eqReviewOwnership}
\end{equation}

\textbf{\cHRev.}
\label{seccHRev}
There is a large literature on review recommendation, \eg~\cite{xia2015should,yu2016reviewer,rahman2016correct}.
%put the citation train back in...
%~\cite{Zanjani2016TSE,balachandran2013reducing,thongtanunam2015should,xia2015should,yu2016reviewer,rahman2016correct,hannebauer2016automatically,lipcak2018large}. 
We note that we did not find a replication package or recommender implementation for any of these works. We re-implement cHRev~\cite{Zanjani2016TSE} because it includes a wide range of factors in its recommendation and has a higher accuracy than the other review recommenders such as RevFinder~\cite{thongtanunam2015should}. 

cHRev scores candidate reviewers by the expertise, frequency, and recency of their past reviews. First, cHRev takes the number of comments made by a candidate on a file as a proxy for expertise. Second, cHRev considers the number of work days a developer has worked on a file as a proxy for measuring effort. Third, cHRev weights recent reviews more highly.

cHRev defines the $xFactor(D, F)$ as the measure of the expertise for a developer $D$ on a file $F$. $C_f$, $W_f$, and $T_f$ respectively show the number of review comments contributed by $D$ for $F$, the number of work days $D$ has dedicated on contributing comments on $F$, and the last day that $D$ worked on $F$. To provide a denominator, $C_{f'}$,$W_{f'}$, and $T_{f'}$ indicate the total number of comments made on $F$, the total number of work days spent on commenting on $F$, and the time of the most recent comment on $F$, respectively.

\begin{equation}
xFactor(D, F) = \frac{C_f}{C_{f'}} + \frac{W_f}{W_{f'}} + \frac{1}{|T_f-T_{f'}|+1}
\end{equation}

To compute the score of a candidate reviewer for a given code review, they sum up the $xFactor(D, F)$ that the candidate, $D$, has on the files in the change, $F$.

%inThesis
%\begin{equation}
%\text{cHRev}(C, R) = \sum_{f}^{Files_{R}} \text{xFactor}(C, F)
%\end{equation}

\subsection{The Turnover Mitigating Recommenders}
\label{secturnovermit}

The focus of existing recommenders on experts disregards the other benefits of code review such as knowledge sharing. Rigby and Bird~\cite{rigby2013convergent} report that code review increases the number of files developers see by between 100\% and 150\%. We speculate that code review can be effective in mitigating the turnover-induced knowledge loss. Based upon this idea, we design reviewer recommenders that spread knowledge to developers with a high retention potential.

\subsubsection{Distributing Knowledge}
\label{secSpreadingKnowledge}

We define a candidate's {\it knowledge} of review request as the number of files under review that a candidate has modified or reviewed in the past divided by the total number of files under review. 

$\text{ReviewerKnows}(D, R) =$
\begin{equation}
 \frac{\text{NumCommitOrReviewedFiles}(D,R)}{\text{NumFilesUnderReview}(R)}
\label{eqKnowledge}
\end{equation}  

Equation~\ref{eqKnowledge}, assigns developers with knowledge of the code under review and ensures expert opinions but concentrates the knowledge of these files exacerbating the risk from turnover. 

\textbf{\learnRec.} To distribute knowledge among the developers, we inverse the $\text{ReviewerKnows}(D, R)$ function to understand how many new files a developer will learn about. We limit the recommender to only display candidates that know about at least one file under review. We then score the remaining reviewers using the \learnRec recommender to maximize learning through the scoring function: 

\begin{equation}
\text{\learnRec}(D, R) = 1 - \text{Knowledge}(D, R)
\label{eqSpread}
\end{equation}

\subsubsection{Developer Retention}
\label{secRetentionKnowledge}

Developers who have made substantial recent contributions to a project have demonstrated a high degree of commitment to the project\cite{constantinou2017empirical,sharma2012examining}. In contrast, assigning a review to a developer who is transient and will likely leave the project is antithetical to the goal of retaining project knowledge. We define commitment and contribution consistency measures to recommend reviewers with a high potential of remaining on the project, \ie high retention potential. In contrast to the previous measures, which are at the pull request or review level, the retention is done at a project-wide level. We use the retention potential for a one-year period, because prior work showed that developers are considered learners up to one year, and that the number of commits they make plateaus at one year~\cite{Zhou2010FSE}. In practice, the development team can adjust this ratio based on their turnover rate. 

\textbf{ContributionRatio.} We measure the contribution of potential of a developer, $D$, by the number of reviews and commits he or she has made in the last year divided by all the commits and reviews on the project.

$\text{ContributionRatio}_{365}(D) =$
\begin{equation}
 \frac{\text{TotalCommitReview}_{365}(D)}{\sum_{d}^{\text{Devs}}\text{TotalCommitReview}_{365}(d)}
\end{equation}

\textbf{ConsistencyRatio.} It is common for developers to make substantial contributions to a feature and leave the project after the feature is complete. To avoid assigning reviews to transient developers, we define the $\text{ConsistencyRatio}_{365}(D)$ as the proportion of months a developer has been active in the last year.

\begin{equation}
\text{ConsistencyRatio}_{365}(D) = \frac{\text{ActiveMonths}_{365}(D)}{12} 
\end{equation}

\textbf{\retentionRec.} We develop \retentionRec that suggests reviewers who are unlikely to leave the project. The scoring function for a candidate review, $D$ is 

$\text{\retentionRec}(D) =$
\begin{equation}
 C1 \cdot \text{ConsistencyRatio}_{365}(D) * C2 \cdot \text{ContributionRatio}_{365}(D)
\label{eqRetention}
\end{equation}

$C1$ and $C2$ are constant coefficients and can weight consistency and contribution. In our simulations, we consider them to be equally important, $C1 = C2 = 1$. In practice, developers could add weights to this equation to emphasize, for example, consistency over total contributions.

\subsubsection{Distribution and Retention Combined}
\textbf{\turnoverRec.} To ensure that knowledge is distributed among developers who are likely to remain on the project, we define the \turnoverRec recommender scoring function for a developer and review as 

\begin{equation}
\label{eq:knowledgeRec}
\turnoverRec(D,R) = C1 \cdot \learnRec(D,R) * C2 \cdot \retentionRec(D) 
\end{equation}

The importance of learning over retention for a project can be adjusted based on the weights, $C1$ and $C2$. In this simulation, we consider learning and retention equally important and set $C1 = C2 = 1$. Project managers can adjust these weights based on their specific goals. 

\textbf{\\ \sofiaEhsan: \turnoverRec and \cHRev combined.}

The \sofiaEhsan recommender distributes knowledge when there are files under review that are abandoned or hoarded, \turnoverRec, and suggests experts, \cHRev, when all files already have multiple knowledgeable developers. The equation shows the \sofiaEhsan scoring function:

$\text{\sofiaEhsan}(D,R) =$
\begin{equation}
\label{eq:sophia}
  \begin{cases}
   \text{\turnoverRec}(D,R), \\ \quad \text{if $|\text{Knowledgeable}(f)| \leq k$, any $f \mid f \in R$} \\
   \text{\cHRev}(D,R), \\ \quad \text{otherwise} 
\end{cases}
\end{equation}

If there are zero knowledgeable people for a file, then the file is an abandoned file, and if there is only one knowledgeable developer for a file, then the file is hoarded by one person. We set $k = 2$ as the threshold for \turnoverRec because this is the smallest number where there is some distribution of knowledge. When there are more than two knowledgeable developers, we recommend an expert with \cHRev. 

\subsection{Workload Aware Recommenders}
\label{secWhoDoMethod}

\textbf{\whodo.} The \whodo recommender was developed at Microsoft by Asthana \etal ~\cite{asthana2019whodo}, and it recommends reviewers based on their workload and recent expertise. The review recommender scoring function is defined as 

\begin{equation}
\label{eqWhodo}
\begin{split}
\text{Score}(D) =   C1 \cdot {\sum_{f \in F}n_\text{change}(D,f)}\frac{1}{t_\text{change}(D,f)} +  \\
C2 \cdot {\sum_{p \in P}n_\text{change}(D,p)}\frac{1}{t_\text{change}(D,p)} +\\  
C3 \cdot {\sum_{f \in F}n_\text{review}(D,f)}\frac{1}{t_\text{review}(D,f)}+\\
C4 \cdot {\sum_{p \in P}n_\text{review}(D,p)} \quad \frac{1}{t_\text{review}(D,p)}\\
\end{split}
\end{equation}

Where $D$ is the candidate reviewer, $f$ is the files in the change, $p$ is the set of last-level parent directories that are changed. The number of commits and reviews that $D$ has done to the file $f$ is  $n_\text{change}(D,f)$ and $n_\text{review}(D,f)$, respectively. $n_\text{change}(D,p)$ is the number of times reviewer $D$ has committed changes within
directory $p$ and $n_\text{review}(D,p)$ is the number of reviews $D$ has performed on files in directory $p$.

$t_\text{change}(D,f)$ and $t_\text{change}(D,p)$ are the number of days since developer $D$ changed a file and parent directory respectively. $t_\text{review}(D,f)$ and $t_\text{review}(D,p)$ are the number of days since a reviewer $D$ reviewed the file or directory, respectively. Developers who have recently modified or reviewed the files in the directory in current change will have a higher WhoDo score.

$C1$, $C2$, $C3$, and $C4$ are constant coefficients and can weight authorship and reviewership. We follow the creators of WhoDo and set them to one to equally consider authorship and reviewership. 

\textbf{Workload balancing.} WhoDo balances the workload of reviewers by weighting the score by the number of active, open reviews that are assigned a candidate reviewer.

\begin{equation}
\text{\text{Load}(r)} =  e^{\theta.TotalOpenReviews} 
\end{equation}

$\theta$ is a parameter between 0 and 1 to control the amount of
load balancing and we follow the authors of WhoDo set it to $0.5$.

WhoDo's final scoring function is
\begin{equation}
\text{\whodo}(r) = \frac{\text{Score}(r)}{\text{Load}(r)}
\end{equation}

%\subsection{Ownership, Workload, and \far Aware Recommendation}
\textbf{\sofiaTwo: \turnoverRec and \whodo combined.}
To make \sofiaEhsan aware of workload, we replace cHRev with WhoDo. Specifically, when we have hoarded or abandoned files in a review we use \turnoverRec to distribute knowledge, otherwise, we use the \whodo recommender to find an expert developer with a low active review workload. The equation shows the \emph{SofiaWorkLoad} scoring function:

$\text{\sofiaTwo}(D,R) =$
\begin{equation}
\label{eq:sophiaTwo}
  \begin{cases}
   \text{\turnoverRec}(D,R), \\ \quad \text{if $|\text{Knowledgeable}(f)| \le k$, any $f \mid f \in R$} \\
   \text{\whodo}(D,R), \\ \quad \text{otherwise} 
\end{cases}
\end{equation}

Following the same rationale fully discussed for Equation~\ref{eq:sophia}, we set $k = 2$ because this is the smallest number where there is some knowledge distribution. We also conduct a sensitivity analysis varying $k$ from 1 to 8 in the threats to validity section. 

Our replication package contains the code implementing each scoring function, the raw and simulated data, and other less promising combinations of recommenders~\cite{replication}.

\subsection{Simulation and Evaluation Measures}
\label{secEvaluationDefinition}
\label{secSimulation}

To evaluate reviewer recommenders, prior works made recommendations for each existing review and compared their result against the actual reviewers who performed the review, \eg~\cite{Zanjani2016TSE,balachandran2013reducing,thongtanunam2015should}.
%
%~\cite{Zanjani2016TSE,balachandran2013reducing,thongtanunam2015should,xia2015should,yu2016reviewer,rahman2016correct,hannebauer2016automatically}. 
%. The developers who actually performed the review are considered to be correct~\cite{Zanjani2016TSE}.
%
For replication purposes, we also compare with the actual reviewers and report the Mean Reciprocal Rank (MRR) for each recommender. MRR is the average of the inverse rank of the highest ranked correct recommendation. For example, if a correct recommendation is on average the third recommendation, the score would be $1/3$. 
%The average of inverse rank across all review recommendations provides the indication of accuracy for the recommender.

A criticism of prior works can be found in Kovalenko~\etal's~\cite{kovalenko2018does} interviews with developers, who state that the recommenders rarely provide additional value because they suggest obvious expert candidate reviewers. This problem is also inherent in the outcome measure, which assumes that the actual reviewers were the best, \ie ``correct" reviewers. 
Kovalenko~\etal~\cite{kovalenko2018does} suggests that we need to account for other perspectives and outcomes beyond simply attempting to predict the actual reviewers. 

To evaluate the impact of reviewer recommendation on diverse outcomes, we perform simulations. Simulation requires us to replace the actual reviewer with a recommended reviewer and to evaluate the outcomes over a period of time. The simulation involves sequentially making recommendations for each review on a project. To train each recommender, we use the entire history prior to the review. The recommenders consider the files under review and, according to the scoring functions defined in Sections~\ref{secOwnership} to \ref{secWhoDoMethod}, they replace one of the actual reviewers with the top recommended reviewer. For example, if DevA actually reviewed the files, but is replaced with top recommended DevB, then the knowledge from the review will be attributed to DevB, not DevA, for future recommendation and for outcome measurement. We only replace one developer to avoid drastically disrupting the peer review process and because Kovalenko~\etal~\cite{kovalenko2018does} showed that developers usually already know at least one expert review candidate.

In our simulations, the goal is to evaluate and compare the recommenders. As a result, we randomly select one reviewer to be replaced for each review. We have over 80k reviews across 5 projects, making it unlikely that this random selection will lead to systematic bias. {\it Each recommender replaces the same reviewer}, eliminating any variation that would come from randomly selecting different reviewers. As a result of selecting the same seeded random reviewer, the variation in outcome measures will only be a result of the recommender. This seeded random approach achieves our goal of evaluating different recommenders, and explicitly does not evaluate how randomly selected reviewers would have impacted the outcome measures. We further discuss this in threats, Section~\ref{secThreats}.

We measure three outcomes: the degree of reviewer \expertise, the \workload of reviewers, and the number of files at risk to turnover, \far. 
%These measures incorporate the reasons interviewed developers conduct review~\cite{bacchelli2013expectations,greiler2016code,sadowski2018modern,asthana2019whodo}. 
We measure the change in the outcomes over the standard quarterly period~\cite{Rigby2016ICSE,nassif2017revisiting}. Each measure is calculated as a percentage change relative to the actual reviewers who performed the review. 
%, for example, expertise of the developer who actually performed the review.
%
For example, if a recommender replaces an expert reviewer with a non-expert ``learner,'' we would expect the measures to report a percentage decrease in both expertise and core workload with a percentage increase in the knowledge distribution of the development team and fewer files at risk to turnover. We define each outcome measure below.

\textbf{\expertise outcome measure.}
Having high expertise ensures having high quality code review~\cite{Fagan1976IBM,bacchelli2013expectations,bosu2015characteristics}. We measure the \expertise for a review as the proportion of files under review that the selected reviewers have modified or reviewed in the past, \ie the union of the files that the reviewers know about. We sum the expertise across the reviews per quarter, $Q$.

\begin{equation}
\text{Expertise}(Q) = \sum_{R}^{Reviews(Q)}{\frac{\text{FilesReviewersKnow(R)}}{\text{FilesUnderReview(R)}}}
\label{eqExpertise}
\end{equation}

\textbf{\workload outcome measure.} Miraseedi and Rigby~\cite{Mirsaeedi2020ICSE} only considered the top 10 core reviewers when measuring the change in workload. This effectively made adding learners outside of the core ``free" because their change in workload was not measured. In contrast, Asthana \etal~\cite{asthana2019whodo} measure the average number of open reviews per developer per day which is appropriate when reviews are spread evenly across developers but averages are misleading when a core group of developers does the vast majority of reviews. 
%However, the average is misleading when done by a small group of core reviewers. 
In our historical analysis section, we show that the top 20\% of reviewers do between 75\% and 84\% of the work, depending on the project. Prior work divided developers into core and non-core developers, however, this dichotomy would require two measures and involves thresholds, \eg the change for the top 20\% vs the bottom 80\% of reviewers~\cite{Mockus2000ICSE, Rigby2016ICSE}. 

\begin{figure}
    \centering
        \includegraphics[width=0.5\textwidth]{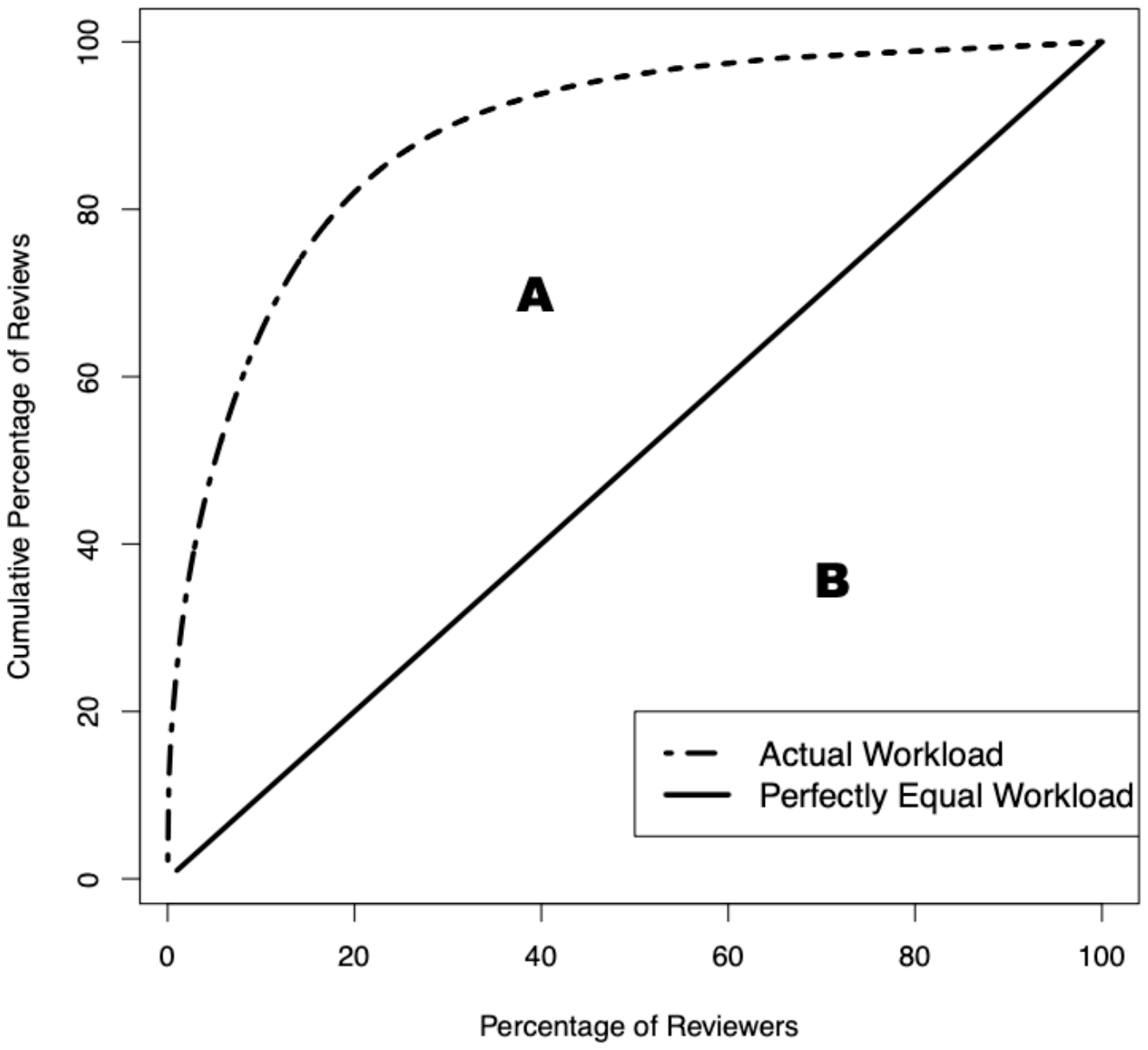}
        \caption{An example illustrating the \workload measure using data from Kubernetes. We see that Actual Workload is highly skewed with the top 20\% of reviewers doing over 80\% of the work. The x = y line represents a Perfectly Equal Workload. The Gini coefficient of workload, \workload, is the area of A divided by the area of B.}
    \label{figWorkloadAUCExample}
    
\end{figure}

To avoid threshold, we adapt the economic measure of the Gini coefficient and the Lorenz curve to understand the concentration of reviewer workload (rather than the concentration of money). The Lorenz curve is a percentage-percentage plot. Figure~\ref{figWorkloadAUCExample} is an inverted Lorenz curve to show the top percentile of reviewers rather than the bottom percentile of reviewers. The $x=y$ line shows an even normal distribution, where 20\% of the reviews are done by 20\% of the reviewers. The curve showing Actual Workload is highly skewed with 20\% of the reviewers doing 82\% of the reviews (see Section~\ref{resultWorkload} for results and discussion). To convert the Lorenz curve to a single workload concentration value, we use the Gini coefficient of workload, \workload. \workload is defined as the area between the Lorenz curve and the perfect equality line divided by the total area under the line of perfect equality. Given the areas A and B shown in the figure, the definition is as follows:

\begin{equation}
\text{\workload}(Q) = A/B 
\label{eqWorkload}
\end{equation}

\textbf{Files at Risk (\far) outcome measure.} We need to quantify the project's exposure to turnover from knowledge loss. Building on Rigby~\etal's~\cite{Rigby2016ICSE} definition of knowledge loss we define the quarterly Files at Risk, \far, as the number of files that are known by zero or one active developers. Given the function $\text{ActiveDevs}(Q, f)$ that returns the set of developers who have modified or reviewed the file, $f$, and have not left the project at the end of the quarter, $Q$, Files at Risk is defined as

\begin{equation}
\far(Q) = \{\,f \mid f \in \text{Files}\ , |\text{ActiveDevs}(Q,f)|\le 1\}
\label{eqFar}
\end{equation}

Files with no active developers are abandoned. Files with only one active developer are at risk to turnover because the knowledge is hoarded by one person. We consider files with two or more active developers to be at lower risk to turnover. 

\textbf{Percentage change in the measures relative to actual values.} The raw outcome measures do not facilitate easy interpretation or comparison. We report the percentage change for a recommender relative to the actual reviewers.
\ifIsThesis 

We use the Equations \ref{eq:dExpertise},\ref{eq:dWorkload}, and \ref{eq:dFar} to report the percentage change of \expertise, \workload, and \far, respectively.

\begin{equation}
\label{eq:dExpertise}
\text{\dExpertise}(Q) =  (\frac{\text{Simulated\expertise}(Q)}{\text{Actual\expertise}(Q)}-1) * 100
\end{equation} 

\begin{equation}
\label{eq:dWorkload}
\text{\dWorkload}(Q) =  (\frac{\text{Simulated\workload}(Q)}{\text{Actual\workload}(Q)}-1) * 100
\end{equation} 

\begin{equation}
\label{eq:dFar}
\text{\dFar}(Q) =  (\frac{\text{Simulated\far}(Q)}{\text{Actual\far}(Q)}-1) * 100
\end{equation} 

\else
Since percentage change is a trivial formula, we illustrate it only for \dWorkload:

%\point{We define the change in each measure for each recommender as}

\begin{equation}
\text{\dWorkload}(Q) =  (\frac{\text{Simulated\workload}(Q)}{\text{Actual\workload}(Q)}-1) * 100
\end{equation} 

\fi

The simulation results for an \textit{ideal reviewer recommender} increases \expertise during review with a positive percentage change in \dExpertise, reduces the skew in the \workload with negative percentage change in \dWorkload, and reduces the number of files at risk, \far, with a negative percentage change in \dFar.

\section{Project Selection and Data}
\label{secData}

We explicitly select well-established large projects with many completed code reviews. On smaller projects, reviewer recommendation is less meaningful as the potential set of reviewers is small and the developers are often aware of the entire team. To select projects, we first query the GitHub torrent dataset to find projects with more than 10K pull requests~\cite{gousios2013ghtorent,replication}. We then apply the following manual selection criteria:

\begin{enumerate}
\item We need existing reviews, so 25\% or more of the commits must be reviewed. 
\item We need to simulate across time, so the project must be 4 or more years old. 
\item We need diverse knowledge and modules, so we ensure there are at least 10K files.
\end{enumerate} 

%\point{the projects we actually picked and the summary stats}
Five projects met our selection criteria. Of these projects, CoreFX, CoreCLR, and Roslyn are led by industry but are available under an open source license and are developed in the open on GitHub. Rust and Kubernetes are community driven OSS projects. Table~\ref{tableSystems} provides summary statistics, including the number of files, pull requests, and commits. Our replication package contains the project data~\cite{replication}.

\begin{table}
  \small
  \centering
  \caption[Systems under study]{Size of projects under study. We explicitly select for large, long-lived projects.}
  \begin{tabular}{l|r|r|r|r}
  \hline
    Project  &  Files  & Reviewed PRs  & Years & Developers \\ \hline
    CoreFX  &  16,015 & 13,499 & 5  & 985 \\ \hline
    CoreCLR  &  15,199 & 10,250 & 4  & 698 \\ \hline
    Roslyn  &  12,313 & 8,646 & 5  & 469 \\ \hline
    Rust  &  12,472 & 17,499 & 9  & 2,720 \\ \hline
    Kubernetes  & 12,792 & 32,400 & 5 & 2,617 \\ \hline
  \end{tabular}
  \label{tableSystems}
\end{table}

\subsection{Gathering Data}

%\point{Raw Data}
We gather authorship commit data from git and review data from GitHub. We clone the repositories to extract all commits and corresponding changes. On GitHub, reviews are conducted in pull-requests that allow the authors and reviewers to discuss each change~\cite{gousios2014exploratory}. 
%\point{who is a reviewer}
In this study, we consider an individual to be a reviewer of a pull-request if he or she writes a review comment on a file, asks for further changes from the author, or approves/rejects the pull request.
%\point{cleaning}
To gather and clean the required data, we developed a post-processing pipeline which we make publicly available~\cite{replication}.

\textbf{Unifying Developer Names.} When a developer makes commits using his or her GitHub username we can link this with the email address they use in the git commit. In some cases, the author commits without using a GitHub username and we use a name unifying approach that employs edit distances to match the git email names with GitHub usernames. This approach is similar to Bird's~\etal's~\cite{bird2006mining} and Canfora \etal's~\cite{canfora2012going}.

%\peter{}
\textbf{Leavers.} Robillard~\etal\cite{robillard2018threats} shows that using the last commit as an indicator for departure of developers draws some risks. Based on this finding, at the end of each quarter, we consider the knowledge of a developer to be inaccessible if he or she has no contribution in the subsequent four quarters. We exclude the last quarter of projects from analysis to ensure that we do not mistakenly label a developer as a leaver if they have gone on vacation for a month more.  

\textbf{Excluding mega commits.} Rigby \etal~\cite{Rigby2016ICSE} argue that commits with hundreds of file changes are too large to be fully comprehended by the author. In manual analysis of mega commits and review requests, we find that they tend to be superficial changes including renaming a folder, renaming a function throughout the source code, changing the license and copyright of files, or importing a large chunk of code from another version control system. We do not associate any knowledge to the author or reviewer of changes with 100 or more files. 

In this work, we also limit our study of knowledge to code files, including \textit{.cs}, \textit{.java}, and \textit{.scala}. We also exclude review comments that are made after the code has been merged, unmerged pull-requests, and files that were committed without review. 

\textbf{Excluding bots.} Finding bots is a semi-manual process. We ranked users by commits, and also searched for users that have ``bot" in their name. In both cases, bots have a significantly higher activity than actual developers. Once, we compiled this list, we manually examined pull-requests comments to ensure they are actually from a bot. GitHub makes it easy to {\it verify} a bot comments as the term ``bot" follows the username, for example, in this comment, ``dotnet-issue-labler bot commented 6 hours ago." The names of each excluded bot across the 80k pull-requests we studied are in the replication package~\cite{replication}.

%Section{Evaluation Setup}

%\todo{define explicitly what you've done: how do the theoretical definitions get applied to the data and produce the outcome results}

\section{Results}
\label{secResults}

In this section, we discuss the results for our research questions relating to (1) a study of knowledge dispersion during review, (2) recommendations based on ownership, including \cHRev, (3) learning and retention aware recommenders, including \sofiaEhsan, (4) a quantification of the workload of reviewers, (5) recommendations based on workload, including \whodo, and a recommender that combines ownership, learning and retention, and workload, \ie \sofiaTwo.  We make three notes. First, we note that RQ1 and RQ4 do not involve simulation and are historical analyses based on the actual reviewers and commits. Second, we note that the MRR outcomes do not involve simulation and instead reports how accurately the recommender predicts the actual reviewers. Third, simulations are run for each recommender, and we note the changes in \dExpertise, \dWorkload, and \dFar as a percentage difference relative to the actual values for each project. Table~\ref{table:average} shows the average for each outcome across all projects. For completeness, the values for each project are in the body of the paper because we would otherwise need 15 tables to display all the results. 

\subsection{RQ1: Review and Turnover (Historical Analysis)}
\label{secTurnoverResults}
%\textit{What is the reduction in files at risk to turnover when both authors and reviewers are considered knowledgeable?}
\textit{What is the reduction in files at risk to turnover when both authors and reviewers are considered knowledgeable of the files in a pull request?}

%\point{prior work ignores review knowledge}
Recent studies have quantified knowledge loss from turnover on the basis of the commits that each developer has made~\cite{Rigby2016ICSE,nassif2017revisiting}. The assumption, in these works, is that knowledge is only attained through writing code. %\point{transfers knowledge}
However, the knowledge transfer that occurs during code review is widely documented with prior work showing that review promotes team awareness, transparency, and shared code ownership, \eg \cite{sadowski2018modern}. 
%\point{attain knowledge}
%Rigby and Bird~\cite{rigby2013convergent} quantified the additional knowledge attained during review and reported that code review exposes developers to between 66\% and 150\% more files than they edit. Thongtanunam \etal~\cite{thongtanunam2016revisiting} found that developers who have not made any changes to a module contributed by reviewing 21\% to 39\% of the code changes in the module.
%\point{remove assumptions}
%In this section, we consider both authors of code as well as reviewers to be knowledgeable and calculate the number of files that are at risk when turnover occurs. 

To assess the extent that the project is at risk to knowledge loss from turnover, we measure \far, see Equation~\ref{eqFar}, which measures the number of files that have zero or one active developers at the end of each quarter. To mirror prior works, we calculate the $\far_{author}$ which only considers authors to be knowledgeable~\cite{Rigby2016ICSE,nassif2017revisiting}. We then calculate \far, which considers both authors and reviewers as knowledgeable. 

%\point{just authors}.
Table~\ref{table:with-without-reviews} reports the proportion of files at risk relative to the total files on the project. The median raw value per quarter of $\far_{author}$ is 7,648, 3,704, 5,602, 2,932, and 5,448 files for CoreFX, CoreCLR, Roslyn, Rust, and Kubernetes, respectively. As a percentage of the codebase, between 68\% and 89\% of the files are at risk of abandonment. 
 %\point{add review}
 In contrast, when both the author and the reviewer are considered knowledgeable, the median raw value per quarter of $\far$ is 1,988, 2,000, 1,918, 1,958, 1,877, respectively. As a percentage of the codebase, between 22\% and 45\% of the files are at risk of abandonment. As a percentage increase in files at risk for $\far$ relative to $\far_{author}$ we see that 74\%, 46\%, 66\%, 33\%, and 65\% fewer files are at risk of abandonment for CoreFX, CoreCLR, Roslyn, Rust, and Kubernetes, respectively. We conclude that considering reviewers to be knowledgeable of the files they review drastically reduces $\far$ and gives a clearer picture of the risk a project is at to turnover than prior works that only considered authors to be knowledgeable~\cite{Rigby2016ICSE,nassif2017revisiting}.

%\point{Necessary?? Compared to the Chrome and Avaya project in}

%\point{Necessary?? Compared to Rigby and Bird}

\begin{table}
\small
\centering
\caption[Impact of reviews on \far]{The proportion of total files that are at risk to turnover. When only authors are considered knowledgeable the proportion of files at risk is drastically higher than when both authors and reviewers are considered knowledgeable.}
\begin{tabular}{l|c|c}
\hline
\far    & Authors & Authors + Reviewers \\ \hline
CoreFX & 89.46\% & 24.74\% \\ \hline
CoreCLR & 86.65\% & 45.56\% \\ \hline
Roslyn & 68.00\% & 22.14\%  \\ \hline
Rust & 78.10\% & 44.51\%\\ \hline
Kubernetes & 76.78\% & 26.04\% \\ \hline \hline
Average & 79.79\% & 32.59\% \\ \hline
\end{tabular}
\label{table:rq1} 
\label{table:with-without-reviews} 
\end{table}

\begin{tcolorbox}
    When only authors are considered knowledgeable an average of 80\% of files are at risk to turnover. When reviewers are also considered knowledgeable, the \far average is 33\%. There is substantial knowledge distribution during code review. 
\end{tcolorbox}

\subsection{RQ2: Ownership Aware (Simulation)}
\label{secOwnershipResults}
%\textit{Does recommending reviewers based on code and review file ownership reduce the number of files at risk to turnover?}
\textit{How does recommending reviewers based on the files they have authored and reviewed in the past impact the number of files at risk to turnover?}

\begin{table*}
\centering
\caption{The average of outcome measures across the projects. MRR is shown for replication purposes. Individual project outcomes are discussed in the paper text. The ideal recommender increases expertise (positive \dExpertise), reduces workload (negative \dWorkload), and reduces files at risk to turnover (negative \dFar).}
\label{table:average}
\begin{tabular}{l|r|r|r|r||l}
\hline
{\bf Recommender} & {\bf MRR} & {\bf \dExpertise} & {\bf \dWorkload} & {\bf \dFar} & {\bf Recommender Scoring Functions} \\ \hline
\authorshipRec & 0.45 & 10.89\%  & -1.74\% & 25.61\% &  Author Commit Ownership\\ \hline
\reviewRec  & 0.37 &  14.56\%  & 7.27\% & 59.65\% &  Review Ownership \\ \hline
%\ownership & 0.45 & 14.83\% & 10.22\% &4.22\% & 57.67\% & Author \& Reviewer\\ \hline
\cHRev & 0.51 & 10.61\% & -1.72\% & 2.15\% & Recent Review Ownership \\ \hline \hline
\learnRec & 0.12 & -41.33\% & -20.24\% & 71.53\% &  Learner: spread knowledge \\ \hline
\retentionRec & 0.34 & 16.37\% & 12.73\% & -18.38\%&  Retention: consistent contributions \\ \hline
\turnoverRec  & 0.19  & -23.62\% &  0.14\% & -29.58\%& Learner and Retention\\ \hline \hline
\sofiaEhsan & 0.45 & 6.86\% & 0.98\% & -28.83\%& (Recent Review Ownership) or (Learner and Retention) \\ \hline \hline
%OwnershipRecency &   & 14.19\% &22.15\% &11.53\%& 36.91\%& Author \& Reviewer Recency (file and directory) \\\hline
\whodo & 0.22  & 7.43\% & -11.27\% & 45.89\%& Recent Ownership and Workload\\ \hline \hline
\sofiaTwo  & 0.17  & 2.96\% & -12.36\% & -27.62\% & (Recent Ownership and Workload) or (Learner and Retention)  \\
\hline
\end{tabular}
\end{table*}

Studies show that teams tend to assign reviews to the owners of files under review~\cite{sadowski2018modern,greiler2016code} and experts who have modified or reviewed the files in the past~\cite{balachandran2013reducing, kovalenko2018does}. In this research question, we run simulations to show how recommending reviewers based on ownership affects project outcome measures.

%\subsubsection{\authorshipRec}
%\point{authorship}
\textbf{\authorshipRec.}
Prior works have adapted developer task recommenders ~\cite{kagdi2008can, mcdonald2000expertise,mockus2002expertise} that use historical authorship data to recommend reviewers ~\cite{Zanjani2016TSE,hannebauer2016automatically}. We partially reproduce these authorship recommendations by using the scoring function defined in Equation~\ref{eqAuthorshipOwnership}.
We use the simulation method described in Section~\ref{secSimulation} and evaluate the impact of \authorshipRec on MRR, \dExpertise, \dWorkload, and \dFar. The average values are shown in Table~\ref{table:average}.

\authorshipRec is successful in predicting the reviewers who actually performed the review with an MRR of 0.53, 0.50, 0.43, 0.46, and 0.37 for CoreFX, CoreCLR, Roslyn, Rust, and Kubernetes, respectively. The average across all projects is 0.45. This means that on average the actual reviewer is ranked 2.04.

From the simulations, we see that assigning reviewers based on their commit ownership, \ie authorship, increases the \expertise in reviews by 7.4\%, 6.3\%, 17\%, 11\%, and 13\%, respectively, with an average of 11\% across the projects. The \workload increases for Rust by 5\%, while it is reduced by -5.7\%, -1.3\%, -5\% and -1.7\% for CoreFx, CoreCLR, Roslyn and Kubernetes, with an average of -2\%.
%The \avgWorkload increases for Rust by 9.35\%, but decreases for the other projects by -10.24\%, -4.62\%, -7.29\%, -7.72\% with an overall average of -4.10\%. 
Although \dExpertise is high for each review, \dFar to turnover has risen across all projects by 30\%, 11\%, 35\%, 36\%, and 16\%, with an average of 26\%.

Developers who have authored the files under review are clearly experts. However, suggesting past authors as reviewers concentrates the knowledge of these files and puts the project at greater risk to turnover as non-authors are not suggested as reviewers. 

%\subsection{Review Ownership (\reviewRec)}
%\point{review ownership}
\textbf{\reviewRec.} The majority of review recommenders have used historical review data, \ie who has reviewed which files or modules in the past, to recommend reviewers~\cite{balachandran2013reducing,jeong2009improving,thongtanunam2015should,xia2015should,yu2016reviewer}. We partially reproduce these review ownership results by using the scoring function defined in Equation~\ref{eqReviewOwnership}.
We use the simulation methodology and outcome measures as described above. 
%described in Section~\ref{secSimulation} and evaluate the impact of \reviewRec on MRR, \dExpertise, \dWorkload, and \dFar show the average outcomes across projects in Table~\ref{table:average}.

\reviewRec is slightly less successful at predicting the reviewers who actually performed the review with an MRR of 0.42, 0.45, 0.38, 0.35, and 0.26 for CoreFX, CoreCLR, Roslyn, Rust, and Kubernetes, respectively. The average across all projects is 0.37, which means the actual reviewer rank is averaged to 2.22.

From the simulations, we see that assigning reviewers based on the files they have reviewed in the past increases review \expertise by 13\%, 10\%, 19\%, 13\%, and 18\% respectively, with an average of 15\% across projects. 
These individuals tend to be top reviewers and we see a corresponding increase in \workload of 2.4\%, 11\%, 4.1\%, 11\%, 7.6\% with an average of 7.2\%. 
%The increase in  \avgWorkload is even more pronounced with past reviewers doing all future reviews the increase is 8.29\%, 24.85\%, 7.29\%, 39.9\%, 21.32\% with an average of 20.33\%.
%
By concentrating recommendations on past reviewers of a file, this recommender has the largest increase in files at risk with \dFar values of 13\%, 31\%, 147\%, 100\%, and 8\%, with an average of 60\%.

%\begin{tcolorbox}
%Recommending reviewers based on the files they have reviewed in the past ensures expertise during review (average increase of 15.17\%), but increases the workload of the top reviewers by on average 20.29\% and differ from the set of actual reviewers with an average MRR of 0.45. Concentrating expertise on the top developers substantially increases the risk of knowledge loss when turnover occurs on average by 65.19\%.
%\end{tcolorbox}

%\subsection{RQ3 \cHRev Recommender}

%\textbf{Does a state-of-the-art recommender reduce the number of files at risk to turnover?}

\textbf{cHRev~\cite{Zanjani2016TSE}.}
\label{seccHRevResults}
This recommender builds upon prior work that leverages information in past reviews~\cite{kagdi2008can}, but also accounts for the number of days a candidate reviewer has worked on a file, and the recency of this work (See Section~\ref{seccHRev} for further details). \cHRev has been show to outperform the other review history based recommenders, including RevFinder~\cite{thongtanunam2016revisiting}. In this research question, we re-implement this state-of-the-art recommender and re-evaluate it. We use the simulation method described in Section~\ref{secSimulation} and evaluate the impact of \cHRev on MRR, \dExpertise, \dWorkload, and \dFar.

In the original cHRev paper, the authors report an average MRR of .67 across four projects~\cite{Zanjani2016TSE}. On our projects, cHRev has an MRR of 0.60, 0.56, 0.50, 0.46, and 0.44, for CoreFX, CoreCLR, Roslyn, Rust, and Kubernetes, respectively. The average is 0.51. This implies that on average the actual reviewer is ranked 1.92. Although the MRR is lower in our reproduction than in the original study, we note that for MRR cHRev outperforms the recommenders we have consider so far.

From the simulations, we see that like the other ownership recommenders, \cHRev increases the \dExpertise in reviews by 10\%, 7.1\%, 13\%, 8.3\%, and 14\%, respectively, with an average of 11\% across projects. \dWorkload decreases by -3.2\%, -0.8\%, -1\%, -0.8\%, -2.8\% with the average of -2\%.
%However, unlike \reviewRec, it reduces the load on average reviewer with \dAvgWorkload of -1.46\%, 3.46\%, 1.71\%, -2.95\%, -5.88\% with an average of -3.09\%. 
\cHRev concentrates knowledge and increases the project's risk to turnover with a \far increase of 6.5\%, 10\%, 10\% in CoreFX, CoreCLR, and Rust, respectively and for Roslyn and Kubernetes the \dFar is reduced at -2.6\% and -14\%, respectively. The average of \dFar across all projects is 2.1\%.

\begin{tcolorbox}
   Ownership aware recommenders concentrate knowledge on a expert authors and reviewers. \cHRev is the most accurate ownership recommender suggesting the actual reviewers with an MRR of 0.51. It increases the degree of \expertise during review by 11\%, with a slight decrease in \workload of 1.7\%.
   %, and a reduction in the \avgWorkload of -3.09\%. 
   However, the files at risk to turnover increase with an average \dFar of 2.1\% as knowledge is concentrated on expert reviewers. 
\end{tcolorbox}

\subsection{RQ3: Turnover Aware (Simulation)}
%\textit{Can we reduce the number of files at risk to turnover by developing learning and retention aware review recommenders?}
\textit{Can we reduce the number of files at risk to turnover by developing learning and retention aware review recommenders?}

%shown that it concentrates
The previous research questions have demonstrated that existing review recommenders concentrate knowledge on experts increasing the risk of knowledge loss from turnover. %obvious experts
%Furthermore, in two large industrial settings, Kovalenko \etal ~\cite{kovalenko2018does} interviewed developers and found that suggesting prior review experts tends to recommend reviewers that are obvious to the author of the change. They state that making obvious recommendations leads to a lack of use of recommenders. They envision a new research path for next generation of recommenders that go beyond suggesting experts. 
In this research question, we investigate how we can mitigate turnover-induced loss and disseminate knowledge using learning and retention measures. 

%\subsection{\learnRec} 
\label{secSpreadingResults}

\textbf{\learnRec.} Without reviewer recommenders, development teams naturally distribute knowledge during review by assigning reviewers who would benefit by learning about the files under review~\cite{sadowski2018modern, bosu2016process, bacchelli2013expectations}. 
Building on this idea, in Section~\ref{secSpreadingKnowledge}, we defined a scoring function that determines how many files a candidate reviewer will learn about. We ensure that the candidate knows at least one of the files that is under review. In this way, we spread knowledge, but ensure that the reviewer has some relevant knowledge. We use the simulation method described in Section~\ref{secSimulation} and evaluate the impact of \learnRec on MRR, \dExpertise, \dWorkload, and \dFar with the average outcomes shown in Table~\ref{table:average}.

As expected, \learnRec does a poor job of predicting the reviewers who actually performed the review with an MRR of 0.17, 0.14, 0.11, 0.10, and 0.09 for CoreFX, CoreCLR, Roslyn, Rust, and Kubernetes, respectively. The average across all projects is 0.12. This means that on average the actual reviewer is ranked 8.3. However, the goal of this recommender was to ensure that developers learn from the review and this shows that it suggests unexpected reviewers.

From the simulations, we see a substantial decrease in \expertise: -42\%, -37\%, -30\%, -53\%, and -45\%, respectively, with an average of -41\% across all projects. 
%The \avgWorkload is drastically reduced as a diverse set of learners are recommended: -28.29\%, -24.85\%, -32.61\%, -37.93\%, and -34.19\% with an average of -31.57\%. 
%CoreWorkload
The \workload is drastically reduced as fewer expert reviewers are assigned reviews: -18\%, -23\%, -22\%, -25\% and -13\% with the average of -20\%. This is the largest decrease of any recommender.
%FaR
Counter-intuitively we see an increase in the files at risk with \dFar values of 27\%, 25\%, 124\%, 110\%, 72\% with an average of 71\%. By selecting non-experts, \learnRec recommends transient developers who are less committed to the project.

%\begin{tcolorbox}
%The recommendations substantially differ from actual reviewers, MRR 0.12. \learnRec substantially reduces \expertise, -35.13\%, but suggests learners reducing the \workload by -39.51\%. Counter-intuitively it makes the project drastically more susceptible to knowledge loss from turnover because it assigns reviews to learners who are less committed to the project, \dFar of 63.04\%.
%\end{tcolorbox}

%\subsection{\retentionRec}
\label{secRetentionResults}

\textbf{\retentionRec.} Assigning reviews to transient developers may distribute knowledge, but does not reduce turnover. In Section~\ref{secRetentionKnowledge}, we define a measure that captures how frequently developers contribute to the project and the number of months in the last year that they are active. We ensure that the candidate knows at least one of the files that is under review. Our goal is to assign reviews to committed developers. We use the same simulation methodology and outcome measures.
%We use the simulation method described in Section~\ref{secSimulation} and evaluate the impact of \retentionRec on MRR, \dExpertise, \dWorkload, and \dFar.

\retentionRec is similar to \reviewRec at predicting the reviewers who actually performed the review with an MRR of 0.55, 0.27, 0.25, 0.41, and 0.22 for CoreFX, CoreCLR, Roslyn, Rust, and Kubernetes, respectively. The average across all projects is 0.34. This implies that on average the actual reviewer is ranked 2.6.

From the simulations, we see an increase in \expertise of 14\%, 12\%, 23\%, 13\%, and 20\%, respectively, with an average of 16\%. These percentages are highest for any recommender outperforming ownership recommenders at ensuring expertise during review. We see a corresponding increase in \workload of  10\%, 18\%, 12\%, 13\%, 10\% 
%and a decrease of  for Kubernetes
with an average of 13\%. However, unlike the ownership and cHRev recommenders, we see a reduction in the files at risk with \dFar values of -27\%, -14\%, -21\%, -6.2\%, and -23\%  with an average of -18\%. \retentionRec selects committed developers who are unlikely to leave the project.

%\begin{tcolorbox}
%\retentionRec is the most successful in ensuring experts, 16.59\%, during review, while reducing the risk of knowledge loss from turnover, -15.91\%. However, by focusing on the most committed developers it also has the greatest increase in \workload, 29.42\%. The MRR of 0.39 indicates that the actual reviewers are more diverse than the recommendations.  
%\end{tcolorbox}

%\subsection{\turnoverRec} 

\textbf{\turnoverRec.} We showed that distributing knowledge through \learnRec does not alleviate knowledge loss and \retentionRec increases the \workload. We combine these approaches to distribute knowledge but to distribute it among individuals who have a higher retention potential. Through Equation~\ref{eq:knowledgeRec}, we defined \turnoverRec that multiplies the learning measure by the retention measure. Again we ensure that each candidate knows about at least one file. We use the same simulation methodology and outcomes. 

\turnoverRec does a poor job of predicting the reviewers who actually performed the review with an MRR of 0.27, 0.18, 0.19, 0.19, and 0.12 for CoreFX, CoreCLR, Roslyn, Rust, and Kubernetes, respectively. The average across all projects is 0.19. This implies that on average the actual reviewer is ranked 5.3.

From the simulations, we see that similar to \learnRec, the \expertise has decreased by -26\%, -25\%, -12\%, -34\%, and -22\%, respectively, with an average of -23\%. However, in terms of \workload there is only a slight increase of 1.5\%, 1.7\% and  1.4\% for CoreFX, CoreCLR and Kuberenest and slight decrease of -1.5\% and -2.4\% for Roslyn and Rust with an average of -0.14\%. The files at risk are reduced with a \dFar of -34\%, -12\%, -41\%, -25\%, and -36\% with an average of -29\%.

\begin{tcolorbox}
\turnoverRec combines learning and retention recommenders and has the greatest reduction in turnover risk, \dFar, -30. However, there is a substantial cost in the reduction of \expertise, -24\%, and a minor increase in \workload, 0.14. The low MRR value of 0.19 indicates that developers naturally focus on reviewers with greater expertise than \turnoverRec.
\end{tcolorbox}

%\subsection{RQ5 \sofia}
\subsection{\sofiaEhsan: Ownership and Turnover Aware}
\label{secResultsSophia}
%\textbf{Can we combine recommenders to balance \expertise, \workload, and \far?}

We have seen that the ownership recommenders, \eg cHRev, can increase expertise, and the combination of learning and retention aware recommender, \turnoverRec, can reduce the files at risk to turnover. We combine these recommenders
%
%\textbf{\sofiaEhsan.}
%\point{not all reviews contain files that are at risk}
%Not all reviews contain files that are at risk of abandonment. As a result, we do not need to distribute knowledge on these files because there is already a sufficient number of developers to mitigate knowledge loss from developer turnover. 
%
in Equation~\ref{eq:sophia}. \sofiaEhsan distributes knowledge during review using \turnoverRec when there are files at risk of abandonment. In contrast, when all the files have active developers, \sofiaEhsan uses the \cHRev scoring function to suggest recent experts. Of the 13,690, 10,256, 10,388, 17,810, and 32,260 reviewed pull request on CoreFX, CoreCLR, Roslyn, Rust, and Kubernetes around 25\%, 26\%, 30\%, 29\%, and 17\%, contain files at risk. The remaining pull requests use \cHRev recommendations to ensure sufficient expertise. We use the simulation method described in Section~\ref{secSimulation} and evaluate the impact of \sofiaEhsan on MRR, \dExpertise, \dWorkload, and \dFar with average outcomes shown in Table~\ref{table:average}.

\sofiaEhsan does a good job of predicting the reviewers who actually performed the review with an MRR of 0.56, 0.50, 0.40, 0.41, and 0.39 for CoreFX, CoreCLR, Roslyn, Rust, and Kubernetes, respectively. The average across all projects is 0.45. This implies that on average the actual reviewer is ranked 2.3.

From the simulations, we see that by only distributing knowledge when files are at risk and otherwise suggesting experts, \sofiaEhsan inherits the best characteristics of \turnoverRec and \cHRev. The \expertise goes up by 5.7\%, 3.9\%, 8.9\%, 5.1\%, and 11\%, respectively, with an average of 6.9\%. In terms of \workload, we see a slight increase of 2.8\%, 1.4\%, 1.7\% and 1.9\% for CoreCLR, Roslyn, Rust and Kubernetes.% nmuber for corefx is 0.07% when rounded is 0 so i didn't know how to write that part.
The average of \dWorkload is a minor increase of 1.6\%. \sofiaEhsan distributes knowledge to developers who have a high retention potential and reduces the risk of turnover as measured by a decrease in \dFar of -32\%, 14\%, -43\%, -20\%, and -35\%, with an average of -28\%.

\begin{tcolorbox}
The \sofiaEhsan recommender distributes knowledge when there are files under review that are at risk of abandonment and suggests experts when all files already have multiple knowledgeable developers. This strategy allows us to increase the level of \expertise during review, 6.9\%, while having a minor impact on \workload, 1.6\%, and substantially reducing the number of files at risk by -29\%. \sofiaEhsan also does a reasonable job of predicting the actual reviewers with an MRR of 0.45.
\end{tcolorbox}

\subsection{RQ4: Review Workload (Historical Analysis)} 
\label{resultActiveReviews}
\label{resultWorkload}
\textit{How is the review workload distributed across developers?}

Except for \learnRec, that distributed knowledge to learners at the great expense of reducing \expertise and increasing \far, none of recommenders we evaluated were able to have a substantial reduction in \workload. 
%Before we present results for \workload aware recommenders, we first quantify the distribution of \workload.
%Recent works that interview reviewers, find that experts tend to be overloaded with their review workloads~\cite{sadowski2018modern,greiler2016code}. 
%motivation
Before we describe recommenders designed to be aware of a candidate reviewer's workload, we want to understand the current distribution of review effort and to replicate the workload measures and results of three prior works~\cite{rigby2014peer,kononenko2016code,asthana2019whodo}.

Rigby \etal~\cite{rigby2014peer} found that on six major open source projects including Apache and Linux, the median reviewer participates in only 2 to 3 reviews per month. In contrast, the top reviewers, \ie those at the 95th percentile, participated in between 13 and 36 reviews, depending on the project. 
%They report that the top reviewer on the Apache project participated in 72\% of all reviews. 
%
Replicating these {\it monthly} results on CoreFX, CoreCLR, Roslyn, Rust, and Kubernetes, we find that the median reviewer participates in 2, 2, 4, 2, and 3 reviews per month, respectively. Top reviewers at the 95th percentile are involved 28, 25, 51, 34, and 37 reviews per month. Our results are consistent with Rigby \etal's~\cite{rigby2014peer} and show the same skew in effort of the median reviewer vs the top reviewers.

%Mozilla
Kononenko \etal~\cite{kononenko2016code} interviewed Mozilla developers and found a similar pattern at the weekly level with Mozilla's top reviewers having between 11 and 20 reviews per week. 
%\todo{check that this is an interview range and not measured from the data}
%
We replicate these {\it weekly} results and find that the median reviewer participates in 2, 1, 2, 2, and 2 reviews per week, with the 95th percentile top reviewers participate in 12, 11, 24, 16, and 20. Although Kononenko \etal interviewed developers, our measured 95th percentile range is 11 to 24, which clearly confirms their interviewed values and validates them on different open source projects.  

%We don't have time information, so probably move to 
%Interviews with developers found that reviews consume a large proportion of a developer's work day with a lower bound of 3.2 hours per day at Google~\cite{sadowski2018modern}, 6.4 hours on 45 open source projects~\cite{bosu2013impact}, and 

At Microsoft, Asthana \etal~\cite{asthana2019whodo} find that the {\it daily} per developer average number of open pull request reviews is between 1.3 and 2.7 on small projects and 8.4 on a large project.
On the projects in our study, we find the average review daily workload is 1.8, 2.6, 2.5, 2.2, and 2.9, respectively. However, the distribution is skewed and find with the that median daily values of 1, 1, 2, 1, and 1. At the 95th percentile, the daily number of reviews is 7, 5, 7, 8, and 8. While the average daily values on our studied projects are comparable with the Microsoft values, unfortunately, Asthana \etal did not publish the median or skew of their data. Future work is necessary to determine if review effort is also skewed in a commercial dataset. 

\begin{figure}
    \centering
        \includegraphics[width=0.5\textwidth]{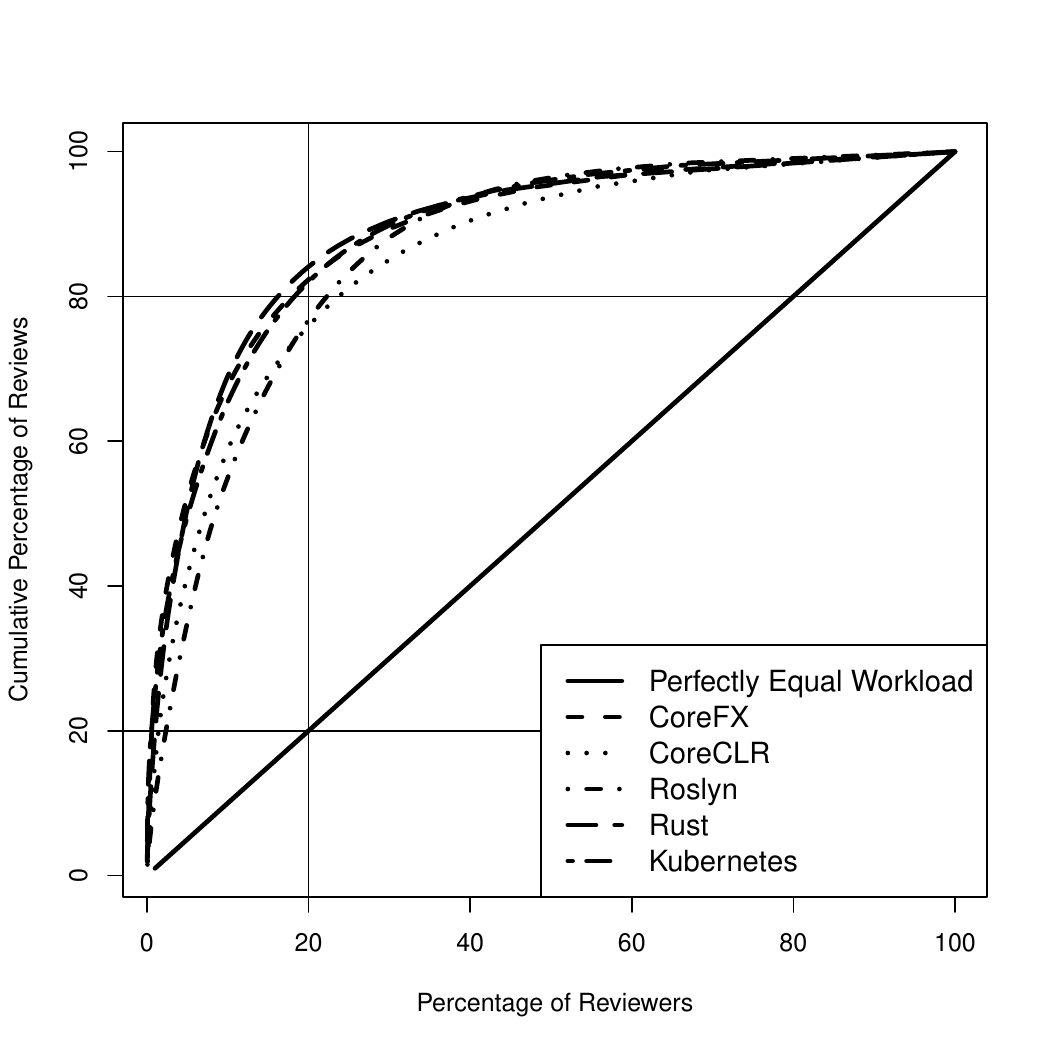}
        \caption{Inverted Lorenz curves of the review workload per quarter. No project has a normal or even distribution where 20\% of the work is done by 20\% of the development team. We see that review effort is highly concentrated with 20\% of the developers performing around 80\% of the reviews.}
    \label{figWorkActual}
\end{figure}

Our replication results show that review effort is not evenly distributed across the open source development teams. We examine the distribution of review effort to further understand the skew.
%In a skewed distribution the average can be misleading. 
Prior works of developer effort examined the commit workload of the top 20\% of committers, \eg~\cite{Mockus2000ICSE,Rigby2016ICSE}. Figure~\ref{figWorkActual} plots the cumulative distribution as the percentage of reviews vs percentage of reviewers. We see that the distributions are highly skewed and that the top 20\% of reviews account for 82\%, 75\%, 76\%, 84\%, and 82\% depending on the project with an average of 80\%. Since the review workload on these open source projects shows a strong Pareto ``80-20 rule," the average workload is misleading. 

\begin{figure}
    \centering
        \includegraphics[width=0.5\textwidth]{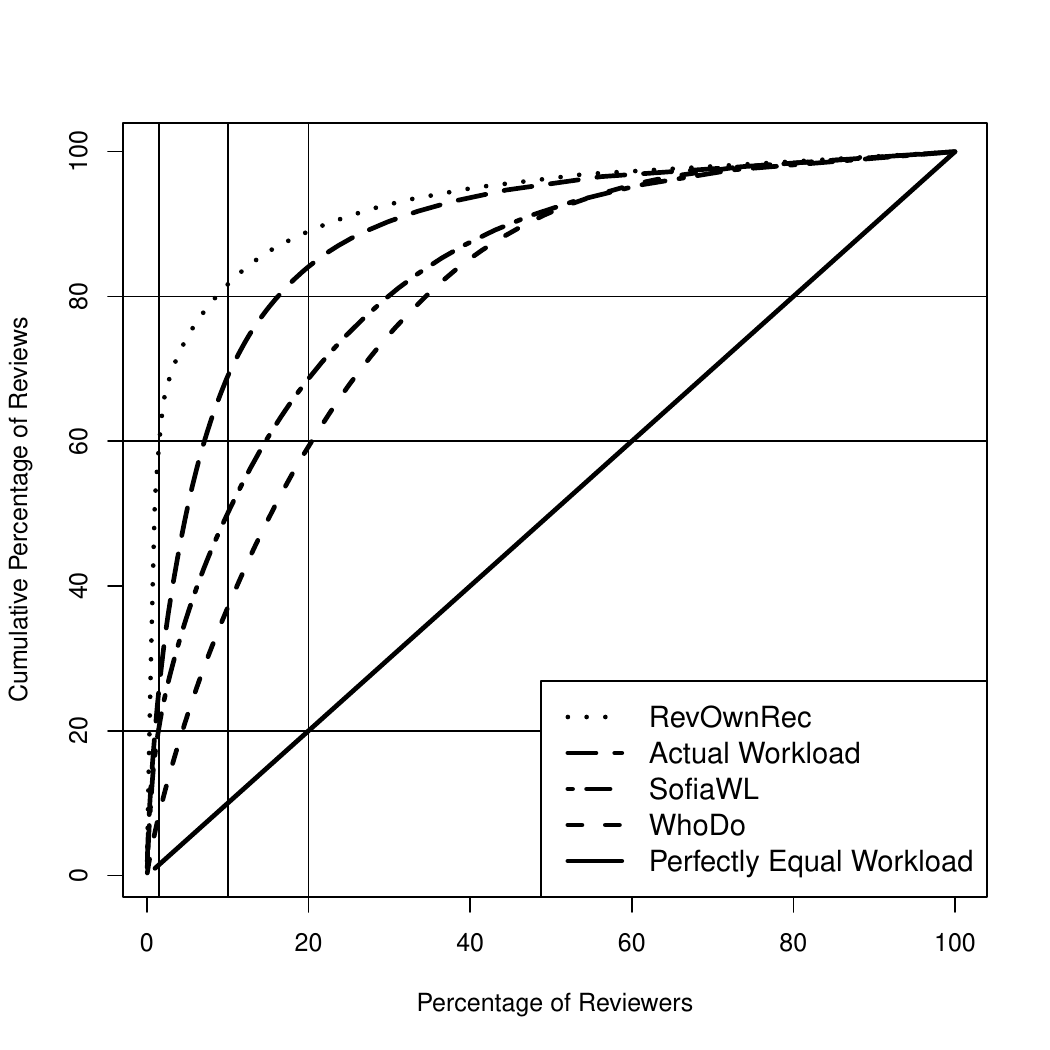}
        \caption{On the Rust project, 20\% of the reviewers do 84\% of the reviews (Actual Workload). We see that by suggesting past reviewers \retentionRec concentrates workload, and 1.5\% of the reviewers do 59\% of the reviews. In contrast, WhoDo distributes workload more evenly, with 20\% of the developers doing 59\% of the reviews. For \sofiaTwo, which also balances expertise and turnover risk, workload at 20\% is 68\%.}
    \label{figWorkloadAUCDiscussion}
    
\end{figure}

As we discussed in Section~\ref{secEvaluationDefinition}, using a percentile cut-off also introduces an arbitrary threshold, and instead we calculate the Gini coefficient, \workload. In the figure, we see that the area under the curve is similar for the actual workload of each project. In our simulations, we calculate the change in area between the simulated and actual workloads, see Equation~\ref{eqWorkload}. As we can see in Figure~\ref{figWorkloadAUCDiscussion}, the workload area varies dramatically depending on the recommender. Understanding the distribution of review effort, we are now in a position to distribute work more evenly, while ensuring that \expertise remains high and that \far is reduced.

%box
\begin{tcolorbox}
The distribution of review effort is highly skewed on open source projects.
We find that the median daily load is between 1 and 2 reviews, while at the 95th percentile the load is 5 to 8 reviews per day. We find that 20\% of the reviewers are responsible for 80\% of the reviews. By comparing the Lorenz curve and \workload coefficient, we  measure the change in area under the curve to acknowledge the skewed distribution of review effort.
\end{tcolorbox}

\subsection{RQ5: Workload Aware (Simulation)}
%\textit{WhoDo is designed to be workload aware, but can it also balance \expertise, \workload, and \far?}
\textit{WhoDo is designed to be workload aware, but what is its impact on reviewer expertise and the files at risk to turnover?}
\label{secWhoDoResults}

%background
In prior work, review recommenders focus on finding experts and do not measure the impact of recommendation on reviewer workload, \eg ~\cite{Zanjani2016TSE,xia2015should,jeong2009improving}.
%~\cite{balachandran2013reducing,thongtanunam2015should,Zanjani2016TSE,xia2015should,jeong2009improving,hannebauer2016automatically}. 
The review recommender recently introduced at Microsoft, WhoDo~\cite{asthana2019whodo}, recommends developer based on the files and directories they have reviewed or committed to in the past. They then weigh the recommender's score by the daily review workload of each candidate, see Section~\ref{secWhoDoMethod}. We re-implement WhoDo and evaluate it in the context of open source projects using the simulation method described in Section~\ref{secSimulation} and MRR, \dExpertise, \dWorkload, and \dFar as outcome measures.

%MRR
WhoDo balances the load of experts resulting in a diverse set of reviewers and low MRR of 0.27, 0.25, 0.21, 0.18, 0.19 for CoreFX, CoreCLR, Roslyn, Rust, and Kubernetes, respectively. The average MRR across projects is 0.22, which means the actual reviewer rank is averaged to 4.5.

%expertise
From the simulations, we see that WhoDo has a substantial increase in \dExpertise for 6.3\%, 4.1\%, 15\%, 4.1\%, and 8.0\% with an average of 7.4\%.
%workload
The design of WhoDo is workload aware and we see a corresponding decrease in \dWorkload of -14\%, -11\%, -0.6\%, -16\% and -15\% with an average across projects of -11.2\%. Roslyn sees the smallest decrease in workload, but it also the smallest project, see Table~\ref{tableSystems}.
The number of files at risk increases dramatically for WhoDo: 45\%, 13\%, 61\%, 49\%, 61\%, with an average of 46\%.
The side-effect of WhoDo recommending developers with a lower daily workload, is that suggested developers have fewer contributions, and are less commitment to the project, which increases turnover and \far.

%box
\begin{tcolorbox}
WhoDo recommends substantially different reviewers with a low MRR of 0.22. However, it is designed to find experts that have a low workload, and it increases \dExpertise, 7.4\%, while substantially reducing \dWorkload, -11\%. Unfortunately, by recommending reviewers with low workload it suggests transient developers and drastically increases \dFar by 45\%.
\end{tcolorbox}

\subsection{RQ6: Ownership, Turnover, and Workload Aware (Simulation)}
%\textit{Can we combine the recommenders to balance \expertise, \workload, and \far?}
\textit{Can we combine the recommenders to balance reviewer expertise, workload, and the files at risk to turnover?}
\label{secSofiaV2Results}

%motivation
The ownership based recommenders are aware of the files that are recently authored or reviewed by developers (\eg cHRev~\cite{Zanjani2016TSE}), WhoDo ~\cite{asthana2019whodo} is aware of developer workload, and \sofiaEhsan~\cite{Mirsaeedi2020ICSE} is aware of files that are potentially at risk to turnover. We combine these recommenders to create a novel recommender that is aware of the ownership, workload, and knowledge distribution. When there are risky files in the pull requests we use \turnoverRec to distribute knowledge, but when no files are at risk to turnover, we use \whodo to find an expert developer with a low workload (see Equation~\ref{eq:sophiaTwo}). We evaluate \sofiaTwo using the simulation method described in Section~\ref{secSimulation} and the outcome measures MRR, \dExpertise, \dWorkload, and \dFar.

%mrr is the same as WhoDo
The MRR for \sofiaTwo on CoreFX, CoreCLR, Roslyn, Rust, and Kubernetes the MRR is 0.23, 0.20, 0.15, 0.13, and 0.14, respectively. The average MRR across projects is 0.17, which means that the actual reviewer is ranked 4.5. Both \whodo and \sofiaTwo have higher MRR than recommenders focussed on learning, \turnoverRec, but lower than the pure ownership recommenders, such as \cHRev.

%increase in expertise
\sofiaTwo inherits the increase in expertise from the ownership aware recommenders with per project increases of 1.6\%, 0.3\%, 4.4\%, 1.8\%, and 6.6\%, and an average across projects of 3.0\%. While the there is still an increase in expertise, it is lower than \sofiaEhsan and \whodo. 

%workload
\sofiaTwo's awareness of \workload leads to a per-project decreases of -9.3\%, -7.2\%, -14\%, -15\%, and -15\%. The across project average is -12\% is substantial, and comparable to \whodo's decrease. Unlike \whodo that is not able to spread workload for Rosyln, \sofiaTwo is able to substantially decrease workload for all projects.

This workload decrease is coupled with an across projects decrease of -27.6\% in \far, with respective project decreases of -37\%, -6.8\%, -43\%, -20\%, and -31\%. While \turnoverRec has a decrease of -29.5\% in \far, \sofiaTwo maintains a strong decrease without the drastic reduction in \expertise.

\begin{tcolorbox}
With \sofiaTwo, \expertise is increased by 3.0\%, \workload is reduced by -12\%, \far is reduced by -28\%. While there are recommenders that can outperform \sofiaTwo on individual attributes, only \sofiaTwo can balance all three simultaneously. 
\end{tcolorbox}

\section{Threats to Validity}
\label{secThreats}
%In this section, We discuss the threats to the validity of our study.
%

\textbf{Generalizability.} We selected large and successful open source software projects that were led by either industry or a community. On smaller projects, there is no need for reviewer recommendation because the list of candidates is small and obvious to all developers. Future work is necessary to validate our results in other development contexts.

\textbf{Construct Validity.} Following prior works on review recommendation~\cite{Zanjani2016TSE,thongtanunam2015should}, ownership~\cite{greiler2016code,fritz2007does,rigby2013convergent}, and turnover~\cite{Rigby2016ICSE,nassif2017revisiting}, we use the source code file as the unit of knowledge. Knowledge is contained in other documents and at other unit levels. We leave these investigations to future work. We have also provided formulas for each of our measures and scoring functions to facilitate replication.

We have considered all files to be of equal importance in this work. Future work could use measures of file churn, complexity, and centrality to factor the importance of a file into the knowledge loss calculation.

The knowledge acquired by a reviewer will be different from the knowledge of the author. The author will usually know more of the details, while an expert reviewer may know more about other modules and dependencies. In this work, we consider both authors and reviewers to be knowledgeable and able to work on the files when turnover occurs. Future work is required to understand the different types and degree of knowledge that authors and reviewers have.

\textbf{Seeded random replacement of one reviewer per pull-request.}

In our simulations, the goal is to directly compare the recommenders. As a result, we randomly select {\it one reviewer} to be replaced for each review. We only replace one developer to avoid drastically disrupting the peer review process and because Kovalenko~\etal~\cite{kovalenko2018does} showed that developers usually already know at least one expert review candidate.

Each recommender replaces the same reviewer, eliminating any variation that would come from randomly selecting different reviewers. As a result of selecting the same seeded random reviewer, the variation in outcome measures will only be a result of the different recommenders. This seeded random approach achieves our goal of evaluating different recommenders, and explicitly does not evaluate how randomly selected reviewers would have impacted the outcome measures. 

Table~\ref{tableSystems} showed that we examine over 80k reviews across 5 projects, making it unlikely that this random selection will lead to systematic bias. Future work could perform thousands of simulations to understand the variation that a Monte Carlo style of simulation would have. Monte Carlo simulations are typically used when an event is very rare~\cite{Rigby2016ICSE}, \eg a banking crisis, and require more processing power than we have available. In contrast, code reviews happen every day, and we would recommend that in practice developers should select the approach that suits their actual needs, \eg workload balancing vs expertise, and adapt it based on how effective the recommender is measured to be in practice. 

%For example, one team might need more knowledge spreading because an important developer is leaving the team, while another might be more focussed on expertise. 

%We do not replace all reviewers, because Kovalenko \etal \cite{kovalenko2018does} showed that the author is already aware of potential expert reviewers and recommenders need only provide novel recommendations.
%As a further check, over a period of four months, we re-ran our top two techniques, \cHRev and \sofia, a minimum of 215 times for each project. 
%
%For \cHRev, we see a change of -0.04, -0.75, -0.52, and	-0.86 percentage points for MRR, \dExpertise, \dWorkload, and \dFar, respectively. The corresponding values for \sofia, are 0.00, -0.10, 2.24, and 	1.21 percentage points, respectively. The results remain consistent with \sofiaEhsan increasing \expertise with a minor increase in \workload, while drastically decreasing \far.

\textbf{Replication and Reproducability.} 
Existing recommenders including ReviewBot~\cite{balachandran2013reducing}, RevFinder~\cite{thongtanunam2016revisiting}, cHRev~\cite{Zanjani2016TSE}, and WhoDo~\cite{asthana2019whodo} do not provide a replication package or source code for their recommenders. As a result, we re-implemented cHRev for comparison because it outperform other state-of-the-art recommenders and WhoDo because it is aware of each reviewer's workload.
%TIE~\cite{xia2015should}, CORRECT~\cite{rahman2016correct}, Comment Network~\cite{yu2016reviewer} against our outcome measures because their source code was not released.
We also implemented simple authorship and ownership recommenders. Comparing each recommender with existing baseline recommenders reduces the threat of internal validity.
We make all of our code and data available for future researchers~\cite{replication}.

\textbf{Sensitivity analysis for \sofiaTwo.}
\sofiaTwo is a bi-modal recommender that suggests experts with low workload when no files are at risk to turnover, with \whodo, and it spreads knowledge with \turnoverRec when there are files at risk. We set $k = 2$ as the threshold for using \turnoverRec because this is the largest number below which files are hoarded or abandoned. We conduct a sensitivity analysis with $k = 1, \ldots, 8$. The average trend across projects for each $k$ is show in Table~\ref{table:sensativityK}. We see \expertise decreases with increase in $k$ because higher $k$ leads to more  \turnoverRec recommendations that spreading knowledge. We see that $k = 2$ is the best point for the \workload because there are more \whodo recommendations that distributes workload. At $k = 1$, knowledge is concentrated on experts with recommendations from  \whodo, afterwards, \dFar remains consistently below -20\%, indicating that additional knowledge spreading above two knowledge people has little impact on the files at risk to abandonment. 

Each of our recommenders is designed with a particular focus. Table~\ref{table:average} that contains the simulation results for each recommender demonstrates the tradeoffs that developers and managers can make for their projects. Managers can tune the scoring functions described in Section~\ref{secBackgroundDefinitions} based on their projects specific needs. 

\begin{table}
\label{table:sensativityK}
\centering
\caption{We vary the minimum number of knowledge developers per file: $k = 1, \ldots, 8$. In the paper, we use $k = 2$ because this is the first time more than 1 developer knows about a file, and it also provides the best balance in outcomes.}
\begin{tabular}{l|c|c|c}
\hline
$k$ & \dExpertise & \dWorkload & \dFar \\ \hline
1  & 6.79\% &-3.18 & -4.45\% \\ \hline
%2 & 3.20\% & -12.00 &  -23.92\%\\ \hline
2 & 2.96\% & -12.36\% & -27.62\%\\ \hline
3 & 2.77\% &-2.29 & -22.02 \%  \\ \hline
4 & 0.39\% & 0.11\% & -23.42\%\\ \hline
5 & -3.78\% &0.40\% &-21.69\% \\ \hline
6 & -4.17\% & -3.75\% & -25.88\%\\ \hline
7 & -7.01\% & 0.68\% & -24.68\%\\ \hline
8 & -7.48\% & 3.14\% &-23.35\%\\ \hline
\end{tabular}
\end{table}

\section{Discussion and Literature}
\label{secRelatedWork}

We position our findings in the research literature. We discuss how we advance our understanding of code review practice, distribute review workload, mitigate turnover risk through \far, and evaluate reviewer recommender systems on diverse outcome measures.

\subsection{Understanding Code Review Practice}
%\point{Motivation for studying code review}
Fagan~\cite{Fagan1976IBM} introduced software inspections in 1976 with a detailed experiment that conclusively showed that inspection found defects earlier in the design process and that unreviewed design artifacts lead to defects that slipped through to the latter stages increased overall effort. In the subsequent 40 years, code review has been extensively studied. Early works focused on examining the process~\cite{Fagan1976IBM,fagan2002design}. However, Porter~\etal~\cite{porter1998understanding} demonstrated that process was much less important than ensuring expertise during review. Current code review practice favors a lightweight process that focuses on expert discussion of changes to the system~\cite{bosu2016process,rigby2011understanding,bacchelli2013expectations,rigby2013convergent,bosu2015characteristics} that still improves software quality~\cite{rigby2014peer,mcintosh2016emse,Krutauz2020EMSE}. 
%\point{We need expertise and we still maintain it during review}
%\peter{}
%*****what we had:*****
%In Section~\ref{secOwnershipResults}, we show that recommenders that focus on past authorship and past review can increase the expertise by up to 19.57\% and 22.12\%, respectively. 
We show that \retentionRec has the highest increase in \expertise among all expert recommenders with an average of 16\%. \reviewRec and \authorshipRec that focus on ownership have an average of 15\% and 11\%, respectively. We also found that focusing on learners will reduce \expertise by up to -27\%.
%Samaneh: which algorithm is this -26.55\% belong to?
%\point{overloaded experts}
\subsection{Review Workload}
\label{secWorkloadDiscussion}

Recent works that interview reviewers, find that experts tend to be overloaded with their review workloads~\cite{sadowski2018modern,greiler2016code} and that it is often difficult to find an available expert reviewer~\cite{greiler2016code,rigby2011understanding,thongtanunam2015should}. Moreover, high overall workload can lead to poor review participation\cite{thongtanunam2017review}, requesting feedback from experts can lead to delays from lack of availability, and can result in fewer opportunities for knowledge dissemination~\cite{greiler2016code}.
%\point{but experts are overworked, so we spread the workload across a larger group of knowledgeable developer.}
%
In Section~\ref{resultWorkload} and Figure~\ref{figWorkActual}, we show that the distribution of review workload is highly skewed with the top 20\% of reviewers performing 80\% of the reviews. While prior works have measured the number of reviews that developers perform over varying time periods~\cite{rigby2014peer,kononenko2016code,asthana2019whodo}, we are the first to demonstrate this strong Pareto principle with review workload. 

Our recommenders show that the workload distribution can be changed without sacrificing \expertise. For space reasons we cannot plot the 45 lines that would represent all 9 recommenders and 5 projects (Please see the figures in our replication package~\cite{replication}). In this discussion and Figure~\ref{figWorkloadAUCDiscussion} we focus on the Rust project.
On the Rust project, 20\% of the developers do 84\% of the reviews. Prior works that recommend reviews solely on the basis of the files that have been reviewed in the past~\cite{Zanjani2016TSE,thongtanunam2016revisiting}, further concentrate the review effort. \reviewRec, which represents this category of recommenders has the top 20\% of reviewers doing 88\% of the reviews. However, the figure clearly shows that the concentration of effort is more extreme with the top 10\% of reviewers doing 81\% and 1.5\% of reviewers doing 59\% of the effort. This concentration of effort is extreme when we consider that 1.5\% of rust developers actually do only 25\% of the work.

In contrast, WhoDo does an excellent job of reducing workload. In the figure we see that the top 20\% of reviewers do 59\% of the reviews. While WhoDo balances expertise, it drastically increases turnover risk. For \sofiaTwo the top 20\% do 68\% of the reviews while maintaining expertise and reducing turnover risk. Clearly there is no direct tradeoff between the outcomes, and development teams can experiment with scoring functions to reduce the review workload on experts and spread knowledge. 

\subsection{Turnover-Induced Knowledge Loss and Mitigation}

%\point{turnover and knowledge loss}

Turnover deprives the project of the leaver's experience and knowledge~\cite{huselid1995impact,ton2008managing} and has been shown to increase the number of defects~\cite{mockus2010organizational}. Previous research has quantified the knowledge loss from turnover and shown that projects with very high turnover are susceptible to as much as five times the expected loss~\cite{Rigby2016ICSE,nassif2017revisiting}. However, these works considered authorship as the only way of gaining knowledge about files. 

%\point{we recalculate knowledge loss}
In contrast with prior work, we include the knowledge gained from conducting reviews into the turnover risk calculations because interviews with developers show that code review is an opportunity for learning and it plays a vital role in distributing knowledge\cite{rigby2013convergent,sadowski2018modern,bacchelli2013expectations,greiler2016code,bosu2016process}. Two separate studies quantified the knowledge gained during code review and showed that at both Google~\cite{sadowski2018modern} and Microsoft~\cite{rigby2013convergent} code review doubles the number of files that developers know. Furthermore, Thongtanunam \etal~\cite{thongtanunam2016revisiting} showed that reviewers of modules are often not authors of the module~\cite{thongtanunam2016revisiting}. 
In Section~\ref{secTurnoverResults}, our empirical results show that review naturally reduces turnover risk. We show that when only authors are considered knowledgeable an average of 79\% of the total files are at risk. When both authors and reviewers are considered knowledgeable the average \far is 32\%. This reduction in \far shows that substantial knowledge is spread during code review.
%However, when we design a recommender that focuses soley on knowledge spreading, we reduce \workload, but counter-intuitively find that far increases dramatically because we recommend candidates who would learn from the review, but who have a low commitment to the project, \ie they are transient. 

%\point{expand turnover studies to include knowledge attained during review, show that it reduces \far} 

%\peter{}
% *******what we had:*******
In this work, we design recommenders that explicitly distribute knowledge by suggesting reviewers who would learn about the files under review. We show that by distributing knowledge among developers who have a higher retention potential, there is a \far reduction of -30\% and -28\% for \turnoverRec and \sofiaTwo, which outperforms both \cHRev and \whodo which increase \far by 2.2\% and 46\%.

Prior works on turnover mitigation suggest increasing documentation with blogs, formalizing the process of documenting bugs in issue trackers, and participating in StackOverflow and internal QA forums~\cite{rashid2017exploring,pee2014mitigating}. Each strategy requires additional developer effort especially for developers who are expert enough to answer questions and write documentation. In contrast, another advantage of using code review in mitigating knowledge loss is that it adds little additional effort because code review is already a common practice on software teams.

\subsection{Recommenders}

%\point{motivation}

Identifying the right reviewers for a given change is a challenging and critical step in the code review process~\cite{Fagan1976IBM,bacchelli2013expectations,greiler2016code,balachandran2013reducing,thongtanunam2015should,Zanjani2016TSE}. Inappropriate selection
of reviewers can slow down the review process~\cite{thongtanunam2015should} or lower the quality of inspection~\cite{bosu2015characteristics,bacchelli2013expectations}. The research on reviewer recommenders focus on the problem of automatically assigning review requests to the expert developers who are most likely to provide better feedback ~\cite{balachandran2013reducing,thongtanunam2015should,Zanjani2016TSE,jeong2009improving,xia2015should,hannebauer2016automatically,yu2016reviewer,Chueshev2020ICSME}.

%\point{Prior works focused on accuracy and it's useless}
Advanced recommenders have been proposed which are built upon machine learning~\cite{jeong2009improving}, text mining~\cite{xia2015should}, social relation graphs~\cite{yu2016reviewer}. However, these papers do not provide public implementation of their recommenders. Re-implementing and testing these recommenders against our outcome measures is beyond the scope we set for this paper. We hope future work will examine these recommenders, and we release all our code and data to facilitate replication and advancement of review recommenders~\cite{replication}.

The existing recommenders have been evaluated using accuracy metrics such as \textit{Top-K} and \textit{MRR} that measure how accurately the recommendations match the actual developers that were involved in a review. This evaluation is based upon the assumption that actual reviewers were among the best candidates to review a change~\cite{kovalenko2018does}. However, it is reported that the focus on accuracy rarely provides additional value for developers because the recommendations are obvious~\cite{kovalenko2018does}. Furthermore, in teams with strong code ownership, finding relevant experts is not problematic~\cite{sadowski2018modern}.
%\point{show why MRR is meaningless}
For replication completeness we calculated MRR. Our results confirm Kovalenko~\etal's findings that a broader perspective is needed when evaluating recommenders. We showed that recommenders with similar MRR values may have entirely different impact on \expertise, \workload, and \far. For instance, \reviewRec and \retentionRec have a difference of 0.03 in MRR while the difference between their \dFar is 78\%. \learnRec and \turnoverRec have a difference of 0.07 in MRR while the difference between their \dWorkload and \dFar is 20\% and 101\%. 

%Samaneh: while the difference between their \dWorkload and \dFar is 40.58\% and 92.58\%.  40.58% is not correct. 

\section{Conclusion}

\label{secConclusion}

In this study, we provide empirical results on the concentration of code review workload and the distribution of knowledge across files, \far. We partially replicate existing work on code review recommendation, fully reproduce two state-of-the-art code review recommenders, \cHRev and \whodo on five open source projects, and propose novel recommenders that are of knowledge distribution. We evaluate the mean reciprocal rank (MRR) for replication purposes, but propose a novel evaluation framework for reviewer recommenders based their impact on \expertise, \workload, and Files at Risk to turnover (\far). The main contributions from our empirical results and simulations are as follows.

\textbf{Ownership Aware Recommenders} concentrate knowledge on a small group of experts. Our findings triangulate Kovalenko \etal's~\cite{kovalenko2018does} interview finding that traditional review recommenders suggest experts who already known by author of the change. We see this in the increase \expertise and high MRR, \eg \cHRev 11\% and 0.51, respectively. We further show that this knowledge concentration increases the risk of turnover with \cHRev increasing \far by 2\% and \retentionRec, that suggests past reviewers, \dFar 65\%. 
%Samaneh: where is 65% is coming from?
We show that code review naturally spreads knowledge and develop recommenders to enhance this learning effect. We develop three novel \textbf{Turnover Aware Recommenders.} The first, \learnRec, suggests the develop who knows only one file that is under review. This recommender reduces \workload but has drastic decreases in \expertise and \far by recommending transient developers. \retentionRec attempts to solve this problem by recommending developers with high retention potential. While effective for both \expertise and \far, it drastically increases core developer \workload. \turnoverRec combines these recommenders, and has the greatest reduction in \far at -29\% but reduces \expertise by -24\%.

\textbf{Workload Distribution.} Prior works counted the number of reviews done by developers~\cite{rigby2014peer,kononenko2016code,asthana2019whodo}. However, we are the first to plot the distribution of reviews across developers and to show a strong Pareto distribution, with the top 20\% of reviewers accounting for aa average across projects of 80\% of code reviews. Given the skewed data, prior measures such as average workload~\cite{asthana2019whodo} or thresholds~\cite{Rigby2016ICSE,Mirsaeedi2020ICSE} are inappropriate. A major contribution of this work is the use of \workload to compare the concentration of review workload under the Lorenz curve.

We re-implement and re-evaluate the state-of-the-art \textbf{Workload Aware Recommender}, \whodo~\cite{asthana2019whodo}, on open source projects. \whodo is able to balance \expertise and reduce \workload by -11\%. Unfortunately, it drastically increases \far by 46\%.

The final outcome of this work is \sofiaTwo that combines the state-of-the-art expert and workload aware recommender, \whodo, with the learning and retention recommender, \turnoverRec. This bi-functional recommender adapts itself to the context of the review. It distributes knowledge when there are files under review that are at risk to turnover, but otherwise suggests experts with low workload. Through simulation we show that \sofiaTwo is the only recommender that balances the three outcomes simultaneously. This strategy allows us to increase the level of \expertise during review by 3\%, reduces the \workload by -12\%, and reduces the number of files at risk with a \dFar of -28\%. Recommenders that are aware of only one or two attributes can outperform \sofiaTwo on individual outcome measures, but only \sofiaTwo is aware of and can balance all three simultaneously. Developers and managers can tune the scoring functions to optimize the review recommendations based on their goals.

This work opens two important avenues of future work. The first, is the use of alternative outcome measures beyond accuracy based measurements for the evaluation of recommenders. We hope that other researchers will, at a minimum, evaluate future recommenders on the basis of expertise, workload, and turnover risk. There are also a wide range of other goals that could be incorporated into recommenders and used as outcomes, for example, the response rate of different types of reviewers. The second, and most important area of future work, is to evaluate these recommenders in practice. We have begun to implement and A/B test variants of \sofiaTwo with an industrial partner. We hope that others will create their own variants. To facilitate this future work, we released our code~\cite{replication} and GitHub bot~\cite{Mirsaeedi2020ICSE} that makes recommendations on any pull-request, so that developers can have turnover aware code review recommendations.

%We release \sofiaEhsan bot as an open source software that fully integrates with GitHub pull requests and provides reviewer recommendations. The recommendations complement a developer's intuition and experience by providing simple rationale for each review candidate, such as showing how active a candidate has been, how many files he or she would learn about if they performed the review, and how many of the files under review they have modified or reviewed in the past. To the best of our knowledge, existing reviewer recommenders including Microsoft's CodeFlow~\cite{rigby2013convergent} and Google's Gerrit~\cite{sadowski2018modern} do not explicitly recommend reviewers based on distributing knowledge to reduce turnover. Future work is necessary to fully evaluate \sofiaEhsan and to understand the costs and benefits of recommending ``learner'' reviewers in practice. 
%\balance
%\pagebreak

%\input{Contents/ConclusionRemarks}

\balance
\bibliographystyle{IEEEtran}
\bibliography{Contents/bibliography}

\end{document}